\documentclass[12pt]{article}
\usepackage{epsfig}
\usepackage{psfrag}
\usepackage{enumitem}
\usepackage{latexsym}
\usepackage{indentfirst}
\usepackage{fancyhdr}
\usepackage{amssymb}
\usepackage{amsmath}
\usepackage{amsfonts}
\usepackage{pifont}
\usepackage{cite}
\usepackage{bbold}
\usepackage{color}
\usepackage[footnotesize]{caption2}
\usepackage{graphicx}
\usepackage[center,footnotesize,hang]{subfigure}
\usepackage{url}
\usepackage{array}

\begin{document}
\begin{titlepage}
\vspace*{-1cm}
\phantom{hep-ph/***}

\hfill{RM3-TH/11-10}

\vskip 2.5cm
\begin{center}
{\Large\bf A Model for Tri-Bimaximal Mixing from a Completely Broken $A_4$}
\end{center}
\vskip 0.2  cm
\vskip 0.5  cm
\begin{center}
{\large Gui-Jun Ding}$^{a,b}$,
{\large Davide Meloni}$^{c}$
\\
\vskip .2cm
$^{a}${\it Department of Modern Physics,}
\\
{\it University of Science and Technology of China, Hefei, Anhui
230026, China}
 \vskip .2cm
$^b${\it Department of Physics, University of
Wisconsin-Madison,}\\
{\it 1150 University Avenue, Madison, WI 53706, USA}\vskip.2cm
$^c${\it Dipartimento di Fisica "E. Amaldi"}
 \\
 {\it Universit\'a degli Studi Roma Tre, Via della Vasca Navale 84, 00146 Roma, Italy}

\end{center}
\vskip 0.7cm
\begin{abstract}
\noindent

We propose a new $A_4$ model in which both the right-handed
neutrinos and right-handed charged leptons transform as $A_4$
singlets. We reproduce tri-bimaximal mixing pattern exactly although
the $A_4$ symmetry is broken completely at leading order in both the
neutrino and charged lepton sectors. The charged lepton mass
hierarchies are controlled by the spontaneous breaking of the flavor
symmetry. The light neutrino spectrum is predicted to be of normal
type and the lightest neutrino is massless at leading order.
Although the reactor angle $\theta_{13}$ is expected to be of order
$\lambda^2_c$ from the next to leading order corrections, this model
cannot be ruled out by current experimental data including the
latest T2K results. Leptogenesis is realized via the resonant
leptogenesis of the second and the third heavy right-handed
neutrinos which are degenerate at leading order. The
phenomenological consequences for lepton flavor violation are
discussed in detail.

\end{abstract}
\end{titlepage}
\setcounter{footnote}{1}
 \vskip2truecm

\section{Introduction}
In the past years, considerable efforts have been devoted to
discrete flavor symmetry and many discrete groups have been
considered as family symmetry groups to derive some mass independent
textures, see Refs. \cite{Altarelli:2010gt,Ishimori:2010au} for
reviews. In particular, it has been realized that tri-bimaximal (TB)
mixing matrix \cite{TBmix}, which is at least a good zeroth order
approximation to the current neutrino oscillation data
\cite{Schwetz:2008er,Fogli:Indication,GonzalezGarcia:2010er,Fogli:2011qn},
can naturally arise as the result of a particular vacuum alignment
of scalars that spontaneously break certain discrete flavor
symmetries. Many discrete groups have been exploited to reproduce TB
mixing so far and the $A_4$ group seems to be especially suitable to
perform this task. It has been demonstrated, through group
theoretical analysis, that the minimal flavor symmetry capable of
yielding the TB mixing without fine tuning is $S_4$
\cite{Lam:2008rs}. However, from the model building point of view,
$A_4$ appears to be the most economical and simplest realization
which naturally produces the TB mixing pattern. There was great
interest in $A_4$ as a family symmetry in the recent past and
various $A_4$ models have been constructed. We can approximately
categorize the $A_4$ models into three classes based on the neutrino
mass generation mechanisms; there exist models in which neutrino
masses arise from higher dimensional effective operators and models
in which neutrino masses are generated via the see-saw mechanisms
(being the canonical type I, type II, type III see-saw mechanisms or
the combination of them  or the inverse and linear see-saw
mechanisms). The third class of $A_4$ models generates neutrino
masses via one-loop or two-loop radiative corrections but they are
rare. Some of the models also try to extend the $A_4$ flavor
symmetry to the quark sector in the framework of the Standard Model
(SM) and in Grand Unified Theories (GUT).
Table \ref{tab:classfication} is an attempt to classify the large
number of $A_4$ models on the market, based on an earlier
classification done in Ref.\cite{Barry:2010zk}. We only list models
where the focus was the study of flavor mixings; however, there are
also papers based on $A_4$ discussing the  dark matter
\cite{Hirsch:2010ru,Haba:2010ag,Esteves:2010sh,Meloni:2010sk} and
the electroweak  constraints/phenomenology
\cite{Toorop:2010kt,Toorop:2010ex,Machado:2010uc,Fukuyama:2010mz,Cao:2011df}.
From Table \ref{tab:classfication}, we can clearly see that the
three lepton doublet fields are assigned as $A_4$ triplet in almost
all the models and much the same happens for the right-handed
neutrino $\nu^{c}_i$  in type I see-saw, for the field $\Delta$ in
type II see-saw and for the $\Sigma$'s in type III (we will call
these fields see-saw fields in the following).
In this paper, we present an $A_4$ model for TB mixing where the right-handed neutrinos $\nu^{c}_i$ transform as
$\mathbf{1}$, $\mathbf{1}'$ and $\mathbf{1}''$ under $A_4$, all the
right-handed charged leptons $e^{c}$, $\mu^{c}$ and $\tau^{c}$ are $A_4$ singlet $\mathbf{1}$ and the three generations of
left-handed lepton
doublets $\ell_i$ are assigned to a triplet $\mathbf{3}$. This assignment has not been considered so far,
as it can be seen from Table \ref{tab:classfication}.
This model is as simple as previous $A_4$ models but the phenomenological predictions are drastically different:
the light neutrino mass spectrum is of normal hierarchy type and the lightest neutrino mass is exactly zero at LO, although three right-handed neutrinos are introduced. The first one does not contribute to the leptonic CP asymmetry even if NLO contributions are included and
leptogenesis is realized via the resonant leptogenesis mechanism of the second and third right-handed heavy neutrinos, which are degenerate at LO. The resulting predictions for lepton flavor violation are distinct from existing $A_4$ models as well. The present model is a complete new
variant of $A_4$ flavor models presented in the literature.

This paper is organized as follows. In Section 2 we discuss the
structure of the model at leading order (LO) and show that the
neutrino mass matrix is exactly diagonalized by TB matrix.
In Section 3 we justify the vacuum alignment assumed in the previous
section by minimizing the scalar potential of the model in the
supersymmetric limit. The next to leading order (NLO) corrections
induced by higher dimensional operators are analyzed in Section 4.
We discuss the phenomenological predictions of the model for
leptogenesis and lepton flavor violation in Section 5. Finally
Section 6 is devoted to our discussions and conclusions. In order to
make the paper self-contained we include Appendix A on the $A_4$
group and its representations.

\begin{table}[hptb]
\begin{center}
\begin{tabular}{|c|c|c|c|c|c|c|c|c|}\hline\hline
   &  $\ell_i$ &  $e^{c}_i$   &  $\nu^{c}_i$ & $\Delta$  &  $\Sigma$ & Quark & GUT &  Refs.\\ \hline
 Effective &    $\mathbf{3}$   & $\mathbf{1}$,$\mathbf{1}'$,$\mathbf{1}''$ & --- & --- & ---  &  \ding{56} & \ding{56} & \cite{Altarelli:2005yp,Zee:2005ut,Altarelli:2005yx,Adhikary:2006wi,Altarelli:2006kg,Honda:2008rs,Brahmachari:2008fn,Feruglio:2008ht,Morisi:2009qa,Chen:2009um,Feruglio:2009iu,Feruglio:2009hu,Barry:2010zk}\\

            & $\mathbf{3}$   & $\mathbf{3}$ & --- & --- & ---  & \ding{56} & \ding{56}   &  \cite{Zee:2005ut} \\

            & $\mathbf{3}$ & $\mathbf{1}$,$\mathbf{1}$,$\mathbf{1}$ & ---& --- & --- & \ding{56} & \ding{56} & \cite{Lin:2008aj} \\

             & $\mathbf{3}$ & $\mathbf{3}$& --- & --- & --- & \ding{52}  & \ding{56} & \cite{Morisi:2011pt}\\

            & $\mathbf{3}$  & $\mathbf{1}$,$\mathbf{1}'$,$\mathbf{1}''$ & ---  & --- & --- & \ding{52} & \ding{56} &  \cite{Bazzocchi:2007na}\\\hline

Type I SS  & $\mathbf{3}$   &
$\mathbf{1}$,$\mathbf{1}'$,$\mathbf{1}''$& $\mathbf{3}$& ---& --- &
\ding{56} & \ding{56} &
\cite{Ma:2001dn,Babu:2002dz,Hirsch:2003dr,Ma:2005qf,Altarelli:2005yx,Yin:2007rv,Adhikary:2008au,Csaki:2008qq,Branco:2009by,Hayakawa:2009va,Bertuzzo:2009im,Hagedorn:2009jy,
Burrows:2009pi,Berger:2009tt,Hagedorn:2009df,Ding:2009gh,Lin:2009sq,Barry:2010zk,Mitra:2009jj,Ahn:2010cc,delAguila:2010vg,Araki:2010ur,King:2011zj}\\

                & $\mathbf{3}$  & $\mathbf{3}$  & $\mathbf{1}$,$\mathbf{1}'$,$\mathbf{1}''$ & ---  &  ---  &  \ding{56}  & \ding{56}  & \cite{Ma:2005mw,Lavoura:2006hb}  \\

                & $\mathbf{3}$  & $\mathbf{3}$ & $\mathbf{3}$ &---  & --- & \ding{56}  & \ding{56} &  \cite{Hirsch:2008rp} \\

                & $\mathbf{3}$ & $\mathbf{1}$,$\mathbf{1}'$,$\mathbf{1}''$ & $\mathbf{1}$,$\mathbf{1}'$,$\mathbf{1}''$ & ---  & --- & \ding{56}  & \ding{56} & \cite{Frampton:2008ci}\\

                & $\mathbf{3}$ & $\mathbf{1}$,$\mathbf{1}$,$\mathbf{1}$  & $\mathbf{3}$ & --- & --- & \ding{56} & \ding{56} & \cite{Hagedorn:2009jy,Altarelli:2009kr,Lin:2009bw,Hagedorn:2009df} \\

                & $\mathbf{3}$ & $\mathbf{1}$,$\mathbf{1}$,$\mathbf{1}$  & $\mathbf{1}$, $\mathbf{1}'$, $\mathbf{1}''$ & --- & --- & \ding{56} & \ding{56} & \cite{Ding:2011gt} \\

                & $\mathbf{3}$   & $\mathbf{1}$,$\mathbf{1}'$,$\mathbf{1}''$& $\mathbf{3}$& ---& --- & \ding{52} &  \ding{56}  & \cite{He:2006dk,Kadosh:2010rm,Kadosh:2011id,Ahn:2011yj}\\

                 & $\mathbf{3}$   & $\mathbf{1}$,$\mathbf{1}'$,$\mathbf{1}''$& $\mathbf{3}$ & ---& --- & \ding{52} & \ding{52}  & \cite{Altarelli:2008bg,Burrows:2010wz}\\

                & $\mathbf{3}$  & $\mathbf{1}$,$\mathbf{1}$,$\mathbf{1}$ & $\mathbf{1}$,$\mathbf{1}$,$\mathbf{1}$ & --- & --- & \ding{52} & \ding{52}  & \cite{King:2006np} \\

                 & $\mathbf{3}$  & $\mathbf{1}$,$\mathbf{1}$,$\mathbf{1}$ & $\mathbf{1}$,$\mathbf{1}$ & --- & --- & \ding{52} & \ding{52}  & \cite{Antusch:2010es,Antusch:2011sx} \\

                 & $\mathbf{3}$  & $\mathbf{3}$ & $\mathbf{3}$ & --- & --- & \ding{52} &  \ding{52}  & \cite{Morisi:2007ft,Grimus:2008tm,Bazzocchi:2008rz,Albaid:2009uv,Albaid:2011vr} \\  \hline

Type II SS & $\mathbf{3}$ & $\mathbf{1}$,$\mathbf{1}'$,$\mathbf{1}''$ & --- & $\mathbf{3}$, $\mathbf{1}$,$\mathbf{1}'$,$\mathbf{1}''$ & ---& \ding{56} & \ding{56}  & \cite{Ma:2004zv,Ma:2005sha,Ma:2011yi}  \\

                & $\mathbf{3}$ & $\mathbf{3}$  & --- & $\mathbf{3}$, $\mathbf{1}$,$\mathbf{1}'$,$\mathbf{1}''$ & ---& \ding{56} &  \ding{56}  & \cite{Hirsch:2005mc}   \\

                & $\mathbf{3}$ & $\mathbf{3}$ &---& $\mathbf{1}$  & ---& \ding{56} &  \ding{56}  & \cite{Ma:2006vq}\\

                 & $\mathbf{3}$ & $\mathbf{3}$ &--- &   $\mathbf{3}$,$\mathbf{1}$  & ---& \ding{56}  & \ding{56}  & \cite{Ma:2009wi}\\

                & $\mathbf{3}$  & $\mathbf{3}$ & -- & $\mathbf{1}$,$\mathbf{1}$ & --- & \ding{52} & \ding{56} &  \cite{Bazzocchi:2007au} \\

                & $\mathbf{3}$  & $\mathbf{3}$ & --  & $\mathbf{3}$, $\mathbf{1}$,$\mathbf{1}'$,$\mathbf{1}''$ & ---& \ding{52} & \ding{52} &\cite{Ma:2006wm}     \\

                & $\mathbf{3}$  & $\mathbf{3}$ & --  & $\mathbf{3}$, $\mathbf{1}'$ & ---& \ding{52} & \ding{52} &\cite{Ciafaloni:2009qs}  \\  \hline

Type III SS & $\mathbf{3}$ &
$\mathbf{1}$,$\mathbf{1}'$,$\mathbf{1}''$ & --- & --- & 3 &
\ding{56} &  \ding{56}  & \cite{Ahn:2011pq}  \\  \hline

Type I+II SS & $\mathbf{3}$ & $\mathbf{3}$ & $\mathbf{3}$ & $\mathbf{1}$ & ---& \ding{56} & \ding{56}  &  \cite{Chen:2005jm}  \\

                  & $\mathbf{1}$,$\mathbf{1}'$,$\mathbf{1}''$ & $\mathbf{3}$ & $\mathbf{3}$ & $\mathbf{1}'$ or $\mathbf{1}''$ & ---& \ding{56} & \ding{56} & \cite{Hirsch:2007kh} \\

            &  $\mathbf{3}$ &  $\mathbf{3}$ &  $\mathbf{1}$,$\mathbf{1}'$,$\mathbf{1}''$& $\mathbf{1}$ & ---& \ding{56} & \ding{56} & \cite{Adulpravitchai:2011rq}\\

                  & $\mathbf{3}$& $\mathbf{1}$,$\mathbf{1}'$,$\mathbf{1}''$ & $\mathbf{3}$ & $\mathbf{3}$,$\mathbf{1}$ & ---& \ding{52} & \ding{56}  & \cite{Dong:2010gk} \\

                  & $\mathbf{3}$& $\mathbf{3}$ & $\mathbf{3}$ & $\mathbf{1}$ & ---& \ding{52} & \ding{52}  & \cite{Bazzocchi:2008sp} \\  \hline

Type I+III SS &  $\mathbf{1}$,$\mathbf{1}'$,$\mathbf{1}''$ & $\mathbf{3}$ & $\mathbf{3}$ & --- & $\mathbf{3}$ & \ding{52}  &  \ding{52} &  \cite{Ciafaloni:2009ub}\\

                   & $\mathbf{3}$ & $\mathbf{3}$  & $\mathbf{1}$,$\mathbf{1}'$,$\mathbf{1}''$ & --- & $\mathbf{1}$,$\mathbf{1}'$,$\mathbf{1}''$ & \ding{52}  & \ding{52} & \cite{Ciafaloni:2009ub}\\

                   & $\mathbf{3}$ & $\mathbf{3}$ & $\mathbf{3}$ & ---& $\mathbf{3}$ & \ding{52} & \ding{52} & \cite{Ciafaloni:2009ub} \\

                   & $\mathbf{3}$ & $\mathbf{1}$,$\mathbf{1}$,$\mathbf{1}$ & $\mathbf{1}$ & ---& $\mathbf{1}$ & \ding{52} & \ding{52} & \cite{Cooper:2010ik} \\  \hline

Inverse SS & $\mathbf{3}$ & $\mathbf{3}$ & $\mathbf{3}$ &---& --- & \ding{56} & \ding{56}  & \cite{Hirsch:2009mx}  \\

                & $\mathbf{3}$ & $\mathbf{3}$ & --- &---& $\mathbf{3}$ & \ding{56} & \ding{56}  & \cite{Ibanez:2009du}  \\  \hline

Linear SS & $\mathbf{3}$ & $\mathbf{3}$ & $\mathbf{3}$  &---& --- & \ding{56} &  \ding{56}  & \cite{Hirsch:2009mx}  \\  \hline

Radiative  & $\mathbf{3}$ & $\mathbf{1}$,$\mathbf{1}'$,$\mathbf{1}''$ & ---& --- & --- & \ding{56} & \ding{56} & \cite{Fukuyama:2010ff}\\  \hline

Only quark  & --- & ---  & --- & ---  & --- & \ding{52} & \ding{56} & \cite{Ma:2002yp,Ma:2006sk,Lavoura:2007dw,Machado:2011gn}  \\\hline\hline

\end{tabular}
\caption{\label{tab:classfication} \it Classification of $A_4$
models in terms of the neutrino mass generation mechanisms and the
transformation properties of the matter fields presented in the
literature. The notations $\ell_i$, $e^{c}_i$ and $\nu^{c}_i$
represent the left-handed lepton doublet, right-handed charged
lepton and right-handed neutrinos, respectively. $\Delta$ denotes
the Higgs triplet in type II see-saw mechanisms, $\Sigma$ denotes
the fermion triplet in type III see-saw mechanisms. "Effective"
means that the neutrino masses are generated via effective
operators, the abbreviation "SS" denotes see-saw mechanisms and
"Radiative" indicates that neutrino masses are induced as one-loop
or two-loop radiative corrections. The symbol \ding{56} and
\ding{52} in the Quark and GUT columns refers to whether $A_4$ has
been extended to the quark sector and embedded into GUT theories,
respectively. For the linear and inverse see-saw neutrino mass
generation \cite{Hirsch:2009mx,Ibanez:2009du}, an additional SM
singlet transforming as $\mathbf{3}$ under $A_4$ is introduced.}
\end{center}
\end{table}

\section{The structure of the model}

\begin{table}[hptb]
\begin{center}
\begin{tabular}{|c|c|c|c|c|c|c|c|c||c|c|c|c||c|c|c|c|}\hline\hline
Fields & $\ell$ & $e^c$ & $\mu^c$  &  $\tau^c$ & $\nu^{c}_1$ &
$\nu^{c}_2$  &  $\nu^{c}_3$ &$h_{u,d}$ & $\varphi$ & $\xi$ & $\phi$
& $\chi$ & $\varphi^{0}$ & $\chi^0$ & $\Delta^{0}$ & $\rho^0$
\\\hline

$A_{4}$ & $\mathbf{3}$ & $\mathbf{1}$ & $\mathbf{1}$ & $\mathbf{1}$
& $\mathbf{1}$  & $\mathbf{1}'$
 & $\mathbf{1}''$ & $\mathbf{1}$ & $\mathbf{3}$ & $\mathbf{1}'$ & $\mathbf{3}$ &  $\mathbf{3}$ & $\mathbf{3}$
& $\mathbf{3}$ & $\mathbf{1}$ &$\mathbf{1}''$ \\\hline

$Z_4$  &  1  & $i$ & -1  & -$i$  & 1 & 1 &1 & 1 & $i$ & $i$  &  1  & 1 &-1 &
1& 1 & -$i$   \\\hline

$Z_2$ &  1  &1  & 1  &1  & -1 & 1 & 1  & 1 & 1 & 1 & -1 & 1 & 1 & 1
& -1 &-1\\\hline\hline
\end{tabular}
\caption{\label{tab:field}\it  The transformation properties of the
matter fields, the electroweak Higgs doublets, the flavon fields and
the driving fields under the flavor symmetry $A_{4}\times Z_4\times
Z_2$.}
\end{center}
\end{table}

In this section, we present the model and discuss the LO results for
lepton masses and flavor mixing. We formulate the model in the
framework of type I see-saw mechanisms \cite{seesaw} and
supersymmetry (SUSY) is introduced to simplify the discussion of the
vacuum alignment. The complete flavor symmetry of the model is
$A_4\times Z_4\times Z_2$. The $Z_4$ symmetry distinguishes the
neutrinos from the charged leptons, and it is responsible for the
mass hierarchies of charged leptons; $Z_2$ further distinguishes the
right-handed neutrinos $\nu^{c}_1$ from $\nu^{c}_2$ and $\nu^{c}_3$.
Moreover, the $Z_4\times Z_2$ symmetry plays an important role in
ensuring the needed vacuum alignment. All the fields of the model
with their transformation properties under the flavor symmetry group
are shown in Table \ref{tab:field}. We assign the three generations
of left-handed lepton doublet $\ell_i$ as $A_4$ triplet
$\mathbf{3}$, while the right-handed charged lepton $e^{c}$,
$\mu^{c}$ and $\tau^{c}$ are all invariant under $A_4$. Inspired by
our previous work on $T_{13}$ flavor symmetry \cite{Ding:2011qt},
the three right-handed neutrinos $\nu^{c}_1$, $\nu^{c}_2$ and
$\nu^{c}_3$ transform as $\mathbf{1}$, $\mathbf{1}'$ and
$\mathbf{1}''$ respectively. It is remarkable that this
transformation property is different from many previous $A_4$ models
where the right-handed neutrinos generally form a triplet. In our
model, all right-handed fields, being singlets of $A_4$, are treated
democratically, a more symmetric situation than previous models. The
flavor symmetry is spontaneously broken by four flavon fields
$\varphi$, $\xi$, $\phi$ and $\chi$. At LO the flavons $\varphi$ and
$\xi$ couple to the charged lepton sector, while $\phi$ and $\chi$
couple to the neutrino sector. For the time being, we assume that
the scalar components of the flavon fields acquire vacuum
expectation values (VEV) according to the following scheme:
\begin{eqnarray}
\nonumber&&\langle\varphi\rangle=(0,v_{\varphi},0),~~~~~~~~~~\langle\xi\rangle=v_{\xi}\\
\label{22}&&\langle\chi\rangle=(v_{\chi},v_{\chi},v_{\chi}),~~~~~~~\langle\phi\rangle=(0,v_{\phi},-v_{\phi})\,.
\end{eqnarray}
We will demonstrate that this particular vacuum alignment is a
natural solution of the scalar potential in Section
\ref{sec:vacuum}. We note that if the auxiliary $Z_2$ symmetry was
absent, it would be enough to introduce one flavon field only to
generate the neutrino masses, thus giving a simpler model.
%
However, all the resulting realizations would predict TB mixing in connection with $m_2=0$ or $m_1=m_3$.
%
%
This is obviously not allowed by neutrino oscillation data and the same remains true even after the
NLO corrections are considered. One of the crucial points of our work is the observation that we need to
introduce at least two flavon fields to break the $A_4$ symmetry in the neutrino sector at LO,
if the right-handed neutrinos are assigned to $A_4$ singlets.

\subsection{Charged lepton}
The charged lepton masses are described by the following superpotential:
\begin{eqnarray}
\nonumber&&w_{\ell}=\frac{y_{\tau}}{\Lambda}\tau^{c}(\ell\varphi)h_d+\frac{y_{\mu_1}}{\Lambda^2}\mu^{c}(\ell\varphi\varphi)h_d
+\frac{y_{\mu_2}}{\Lambda^2}\mu^{c}(\ell\varphi)''\xi
h_d+\frac{y_{e_1}}{\Lambda^3}e^{c}(\ell\varphi)(\varphi\varphi)h_d+
\frac{y_{e_2}}{\Lambda^3}e^{c}(\ell\varphi)'(\varphi\varphi)''\\
\nonumber&&~~+\frac{y_{e_3}}{\Lambda^3}e^{c}(\ell\varphi)''(\varphi\varphi)'+\frac{y_{e4}}{\Lambda^3}e^{c}((\ell\varphi)_{3_S}(\varphi\varphi)_{3_S})h_d
+\frac{y_{e_5}}{\Lambda^3}e^{c}((\ell\varphi)_{\mathbf{3}_A}(\varphi\varphi)_{\mathbf{3}_S})+\frac{y_{e_6}}{\Lambda^3}e^{c}(\ell\varphi\varphi)''\xi
h_d\\
\label{10}&&~~+\frac{y_{e_7}}{\Lambda^3}e^{c}(\ell\varphi)'\xi^2h_d+...
\end{eqnarray}
where dots represent the higher dimensional operators which will be
discussed later. Due to the constraint of the $Z_4$ symmetry, the
electron, muon and tauon mass terms appear at different orders in the
expansion in terms of $1/\Lambda$. After electroweak and $A_4$ symmetry breaking,
%
the superpotential $w_{\ell}$ gives rise to a diagonal charged lepton mass matrix:
\begin{equation}
\label{11}m_{\ell}=\left(\begin{array}{ccc}
(y_{e_2}-2y_{e_4}+2y_{e_5})\frac{v^2_{\varphi}}{\Lambda^2}+2y_{e_6}\frac{v_{\xi}v_{\varphi}}{\Lambda^2}+y_{e_7}\frac{v^2_{\xi}}{\Lambda^2}
& 0 &0\\
0&2y_{\mu_1}\frac{v_{\varphi}}{\Lambda}+y_{\mu_2}\frac{v_{\xi}}{\Lambda}&0\\
0&0&y_{\tau}
\end{array}\right)\frac{v_{\varphi}}{\Lambda}v_d\,,
\end{equation}
where $\langle h_d\rangle=v_d$. We see that the electron, muon and
tauon masses are controlled by the first, second and third power of
$v_{\varphi}/\Lambda$ and $v_{\xi}/\Lambda$. Therefore the mass
hierarchies of the charged leptons are naturally recovered if
$v_{\varphi}/\Lambda$ and $v_{\xi}/\Lambda$ are of order
$\lambda^2_c$ \cite{Lin:2008aj,Altarelli:2009kr}, where
$\lambda_c\simeq0.22$ is the Cabibbo angle. We note that the charged
lepton mass hierarchies are determined by the flavor symmetry itself
without invoking a Froggatt-Nielsen mechanism. As it can be seen
from Eq.(\ref{10}), the $A_4$ group in the charged lepton sector is
completely broken by the VEVs of the flavons $\varphi$ and $\xi$ at
LO, since
$T\langle\varphi\rangle=\omega^2\langle\varphi\rangle$ and $T\langle\xi\rangle=\omega^2\langle\xi\rangle$.
However, the lepton flavor mixing is associated with the combination $\overline{m}_{\ell}=m^{\dagger}_{\ell}m_{\ell}$,
which is obviously invariant under $T$, i.e., $T^{\dagger}\overline{m}_{\ell}T=\overline{m}_{\ell}$. Consequently there is still a remnant $Z_3$
symmetry generated by $T$ in the charged lepton mass matrix.

\subsection{Neutrino}
Neutrino masses are generated by type I see-saw mechanism. The superpotential for the neutrino sector is:
\begin{equation}
\label{12}w_{\nu}=\frac{y_{\nu_1}}{\Lambda}\nu^{c}_1(\ell\phi)h_u+\frac{y_{\nu_2}}{\Lambda}\nu^{c}_2(\ell\chi)''h_u
+\frac{y_{\nu_3}}{\Lambda}\nu^{c}_3(\ell\chi)'h_u+\frac{1}{2}M\nu^{c}_1\nu^{c}_1+\frac{1}{2}M'(\nu^{c}_2\nu^{c}_3+\nu^{c}_3\nu^{c}_2)\,.
\end{equation}
The first three terms contribute to the Dirac mass terms whereas the
last two are the Majorana mass terms for the right-handed neutrinos.
One can always set the masses $M$ and $M'$ to be real and
positive by performing global phase transformations of the right-handed
neutrino fields. After electroweak and $A_4$ symmetry breaking, we
obtain the following LO contributions to the Dirac and
Majorana mass matrices:
\begin{equation}
\label{13}m_D=\frac{v_u}{\Lambda}\left(\begin{array}{ccc}
0 & -y_{\nu_1}v_{\phi} & y_{\nu_1}v_{\phi}\\
y_{\nu_2}v_{\chi} & y_{\nu_2}v_{\chi} & y_{\nu_2}v_{\chi}\\
y_{\nu_3}v_{\chi}  & y_{\nu_3}v_{\chi} & y_{\nu_3}v_{\chi}
\end{array}\right),~~~m_{M}=\left(\begin{array}{ccc}
M &0 &0\\
0&0& M'\\
0& M' &0
\end{array}\right)\,.
\end{equation}
The heavy right-handed neutrino mass matrix $m_M$ can be
diagonalized by a unitary transformation
\begin{equation}
\label{14}U_R^{T}m_MU_R={\rm diag}(M,M',M')\,.
\end{equation}
Two of the right-handed neutrinos are degenerate with mass equal to
$M'$ so that the unitary matrix $U_R$ cannot be fixed uniquely; it
can be expressed as:
\begin{equation}
\label{15}U_R=\left(\begin{array}{ccc} 1&0&0\\
0& e^{i\vartheta}/\sqrt{2}& -ie^{i\vartheta}/\sqrt{2}\\
0&e^{-i\vartheta}/\sqrt{2} & ie^{-i\vartheta}/\sqrt{2}
\end{array}\right)\,,
\end{equation}
where $\vartheta$ is an arbitrary phase parameter.
The light neutrino mass
matrix is given by the see-saw formula:
\begin{equation}
\label{16}m_{\nu}=-m^{T}_Dm^{-1}_Mm_D=\left(\begin{array}{ccc}
a&a&a\\
a&a+b&a-b\\
a&a-b&a+b
\end{array}\right)\frac{v^2_u}{\Lambda}\,,
\end{equation}
where
\begin{equation}
\label{17}a=-2y_{\nu_2}y_{\nu_3}\frac{v^2_{\chi}}{\Lambda
M'},~~~~~b=-y^2_{\nu_1}\frac{v^2_{\phi}}{\Lambda M}\,,
\end{equation}
and it is exactly diagonalized by
the TB mixing matrix $U_{TB}$:
\begin{equation}
\label{18}U_{TB}=\left(\begin{array}{ccc}\sqrt{\frac{2}{3}}
&\frac{1}{\sqrt{3}}  &0\\
-\frac{1}{\sqrt{6}}  & \frac{1}{\sqrt{3}}  & \frac{1}{\sqrt{2}}\\
-\frac{1}{\sqrt{6}}  &  \frac{1}{\sqrt{3}} &-\frac{1}{\sqrt{2}}
\end{array}\right)\,.
\end{equation}
In unit of $v^2_u/\Lambda$, the light neutrino masses are given by:
\begin{equation}
\label{19}m_1=0,~~~m_2=3a,~~~m_3=2b\,.
\end{equation}
It is remarkable that the mass of the lightest neutrino is zero,
although three (instead of two) right-handed neutrinos have been considered. As a result, the light neutrino mass spectrum is
predicted to be of normal type. This is a special feature of the present model since generally a massless neutrino is realized with the help
of minimal see-saw mechanisms \cite{Frampton:2002qc}.
Recalling that two observables related to the neutrino mass spectrum have been measured
\cite{Schwetz:2008er,Fogli:Indication,GonzalezGarcia:2010er}:
\begin{eqnarray}
\nonumber&&\Delta m^2_{sol}\equiv
m^2_2-m^2_1=(7.59^{+0.20}_{-0.18})\times 10^{-5}{\rm eV^2}\\
\label{20}&&\Delta m^2_{atm}\equiv m^2_3-m^2_1=(2.45\pm0.09)\times
10^{-3}{\rm eV^2}\,,
\end{eqnarray}
we have $m_2\simeq0.0087{\rm ~eV}$ and $m_3\simeq0.049{\rm ~eV}$.
Then the sum of the light neutrino masses is
$m_1+m_2+m_3\simeq0.0577{\rm ~eV}$. Moreover, we can
straightforwardly obtain the effective mass parameter $|m_{ee}|$ for
the neutrinoless double-$\beta$ decay:
\begin{equation}
\label{21}|m_{ee}|\equiv\Big|\sum_i(U_{PMNS})^2_{ei}m_i\Big|=\frac{m_2}{3}\simeq0.0029\,{\rm
eV}\,,
\end{equation}
where $U_{PMNS}\equiv U_{TB}$ in our case. Therefore the effective
mass $|m_{ee}|$ is predicted to be far below the sensitivities of
the planned neutrinoless double-$\beta$ decay experiments. It is
important to note that the alignment direction of
$\langle\chi\rangle$ preserves a $Z_2$ subgroup generated by the
element $S$. However, an overall negative sign appears if we act on
$\langle\phi\rangle$ with $S$, thus also the residual $Z_2$ symmetry
is broken. As a result, the $A_4$ group is broken to nothing also in
the neutrino sector at LO. However, it can be checked that $m_{\nu}$
is invariant under $S$ and, therefore, an accidental $Z_2$ symmetry
generated by $S$ is still preserved in the light neutrino mass
matrix (in fact, $m_\nu$ is basically determined by
$\langle\phi\rangle^2$ which leaves the $S$ generator invariant). In
summary, the $A_4$ flavor symmetry is broken completely at LO in
both charged lepton and neutrino sectors, nevertheless there is an
accidental $Z_3 \times Z_2$ symmetry in the charged lepton and
neutrino mass matrices, respectively, that ensures diagonal charged
leptons and TB mixing matrix.

\section{\label{sec:vacuum}Vacuum alignment}

The vacuum alignment problem can be solved by the
supersymmetric driving field method introduced in
Ref.\cite{Altarelli:2005yx}. This approach introduces a global
continuous $U(1)_R$ symmetry which contains the discrete $R-$parity
as a subgroup. The flavon and Higgs fields are uncharged under
$U(1)_R$, the matter fields have $R=1$ and the so-called driving
fields $\varphi^{0}$, $\chi^{0}$, $\Delta^{0}$ and $\rho^{0}$ carry
two units of $R$ charge. The LO driving superpotential $w_d$, which is
linear in the driving fields and invariant under the flavor symmetry
$A_4\times Z_4\times Z_2$, is given by:
\begin{equation}
\label{1}w_d=f_1(\varphi^{0}\varphi\varphi)+f_2(\varphi^{0}\varphi)''\xi+M(\chi^{0}\chi)+g_1(\chi^{0}\chi\chi)
+g_2(\chi^{0}\phi\phi)+g_3\Delta^{0}(\phi\chi)+g_4\rho^{0}(\varphi\phi)'\,.
\end{equation}
In the SUSY limit, the equations for the minimum of the scalar
potential are obtained by deriving $w_d$ with respect to each
component of the driving fields. The vacuum structure of the flavons
$\varphi$ and $\xi$ is determined by
\begin{eqnarray}
\nonumber&&\frac{\partial
w_d}{\partial\varphi^{0}_1}=2f_1(\varphi^2_1-\varphi_2\varphi_3)+f_2\varphi_3\xi=0\\
\nonumber&&\frac{\partial
w_d}{\partial\varphi^{0}_2}=2f_1(\varphi^2_2-\varphi_1\varphi_3)+f_2\varphi_2\xi=0\\
\label{2}&&\frac{\partial
w_d}{\partial\varphi^{0}_3}=2f_1(\varphi^2_3-\varphi_1\varphi_2)+f_2\varphi_1\xi=0\,.
\end{eqnarray}
This set of equations admit two un-equivalent solutions, the first one is
\begin{equation}
\langle\varphi\rangle=(v_{\varphi},v_{\varphi},v_{\varphi}),~~~~\langle\xi\rangle=0\,,
\end{equation}
where $v_{\varphi}$ is undetermined. The second solution is:
\begin{equation}
\label{3}\langle\varphi\rangle=(0,v_{\varphi},0),~~~\langle\xi\rangle=v_{\xi}\,,
\end{equation}
with the condition:
\begin{equation}
\label{4}v_{\varphi}=-\frac{f_2}{2f_1}v_{\xi},~~~v_{\xi}~{\rm
undetermined}\,.
\end{equation}
The VEVs $v_{\varphi}$ and $v_{\xi}$ are naturally of the same
order of magnitude (without fine tuning among the parameters $f_1$ and $f_2$), this is consistent with the conclusions drew from
the charged lepton mass hierarchies. Only the second alignment can provide the results of the previous section but we need of some
soft masses in order to discriminate it as the lowest minimum of the scalar potential (not discussed here).
The minimization equations for the vacuum configuration of $\phi$ and $\chi$ are given by:
\begin{subequations}
\begin{eqnarray}
\label{5a}&&\frac{\partial
w_d}{\partial\chi^{0}_1}=M\chi_1+2g_1(\chi^2_1-\chi_2\chi_3)+2g_2(\phi^2_1-\phi_2\phi_3)=0\\
\label{5b}&&\frac{\partial
w_d}{\partial\chi^{0}_2}=M\chi_3+2g_1(\chi^2_2-\chi_1\chi_3)+2g_2(\phi^2_2-\phi_1\phi_3)=0\\
\label{5c}&&\frac{\partial
w_d}{\partial\chi^{0}_3}=M\chi_2+2g_1(\chi^2_3-\chi_1\chi_2)+2g_2(\phi^2_3-\phi_1\phi_2)=0
\end{eqnarray}
\end{subequations}
\vskip-0.15in
\begin{equation}
\label{6}\hskip-0.9in\frac{\partial
w_d}{\partial\Delta^{0}}=g_3(\phi_1\chi_1+\phi_2\chi_3+\phi_3\chi_2)=0
\end{equation}
\begin{equation}
\label{7}\hskip-0.9in\frac{\partial
w_d}{\partial\rho^{0}}=g_4(\varphi_3\phi_3+\varphi_1\phi_2+\varphi_2\phi_1)=0\,.
\end{equation}
Taking into account the alignment of $\varphi$ in Eq.(\ref{3}), we
can infer from Eq.(\ref{7})
\begin{equation}
\label{8}\langle\phi_1\rangle=0\,.
\end{equation}
Then Eqs.(\ref{5a}-\ref{5c}), Eq.(\ref{6}) and Eq.(\ref{7}) admit
the non-trivial vacuum configuration:
\begin{equation}
\label{9}\langle\chi\rangle=(v_{\chi},v_{\chi},v_{\chi}),~~~~\langle\phi\rangle=(0,v_{\phi},-v_{\phi}),~~~v^2_{\phi}=-\frac{Mv_{\chi}}{2g_2}\,,
\end{equation}
with $v_{\chi}$ undetermined.
As we will show in Section \ref{sec:NLO}, all the three lepton
mixing angles receive corrections of order $v_{\chi}/\Lambda$ or
$v_{\phi}/\Lambda$. The solar neutrino angle $\theta_{12}$ is the
most precisely measured one, the experimental departure from its TB
value being at most of order $\lambda^2_c$. Therefore we expect
$v_{\chi}/\Lambda$ and $v_{\phi}/\Lambda$ of the same order of
magnitude, $\sim \lambda^2_c$ as well. As a consequence, the
following relations among the VEVs hold:
\begin{equation}
\label{23}\frac{v_{\varphi}}{\Lambda}\sim\frac{v_{\xi}}{\Lambda}\sim\frac{v_{\chi}}{\Lambda}\sim\frac{v_{\phi}}{\Lambda}\sim\lambda^2_c\,.
\end{equation}
Henceforth we will parameterize the ratio ${\rm VEV}/\Lambda$ by the
parameter $\varepsilon$. Given the symmetry of the superpotential
$w_d$, we can generate other minima of the scalar potential by
acting on the configuration of Eq.(\ref{3}) and Eq.(\ref{9}) with
the element of the flavor symmetry group $A_4$. However, these new
minima are physically equivalent to the original one, they all lead
to the same physics, i.e., lepton masses and flavor mixings, and the
different scenarios are related by field redefinitions. Without loss
of generality, we can analyze the model by choosing the
vacuum in Eq.(\ref{3}) and Eq.(\ref{9}) as the local minimum.

\section{\label{sec:NLO}Next to leading order corrections}
It is important to check that the NLO contributions do not modify too much the successful LO predictions
and that the deviations from TB mixing lie in the experimentally
allowed range. The NLO corrections are indicated by the subleading higher
dimensional operators in the $1/\Lambda$ expansion, which are
compatible with all the symmetries of the model. In the following,
we will study the NLO corrections to the vacuum alignment, to the charged
lepton and to neutrino mass matrices.

\subsection{NLO corrections to the vacuum alignment}

After including the NLO operators the superpotential $w_d$,
depending on the driving fields $\varphi^{0}$, $\chi^{0}$,
$\Delta^{0}$ and $\rho^{0}$, is modified to:
\begin{equation}
\label{24}w_d=w^{0}_d+\delta w_d\,,
\end{equation}
where $w^{0}_d$ is given by Eq.(\ref{1}) and $\delta w_d$ denotes
the NLO terms, suppressed by one additional power of
$1/\Lambda$ with respect to $w^{0}_d$. The correction terms included in $\delta
w_d$ consist of the most general quartic, $A_4\times Z_4\times Z_2$
invariant polynomial linear in the driving fields,
obtained inserting an additional flavon field in the LO terms.
Concretely, $\delta w_d$ is given by:
\begin{equation}
\label{25}\delta w_d=\frac{1}{\Lambda}\Big(\sum^{8}_{i=1}v_i{\cal
I}^{\varphi^{0}}_i+\sum^{10}_{i=1}c_i{\cal
I}^{\chi^{0}}_i+\sum^{2}_{i=1}d_i{\cal
I}^{\Delta^{0}}_i+\sum^{3}_{i=1}r_i{\cal I}^{\rho^{0}}_i\Big)\,,
\end{equation}
where $v_i$, $c_i$, $d_i$ and $r_i$ are complex coefficients with
absolute value of ${\cal O}(1)$; ${\cal I}^{\varphi^{0}}_i$, ${\cal
I}^{\chi^{0}}_i$, ${\cal I}^{\Delta^{0}}_i$ and ${\cal
I}^{\rho^{0}}_i$ denote a basis of independent quartic invariants:
\begin{eqnarray}
\nonumber&&{\cal
I}^{\varphi^{0}}_1=(\varphi^{0}\chi)(\varphi\varphi),~~~{\cal
I}^{\varphi^{0}}_2=(\varphi^{0}\chi)'(\varphi\varphi)'',~~~{\cal
I}^{\varphi^{0}}_3=(\varphi^{0}\chi)''(\varphi\varphi)'\\
\nonumber&&{\cal
I}^{\varphi^{0}}_4=((\varphi^{0}\chi)_{\mathbf{3}_S}(\varphi\varphi)_{\mathbf{3}_S}),~~~{\cal
I}^{\varphi^{0}}_5=((\varphi^{0}\chi)_{\mathbf{3}_A}(\varphi\varphi)_{\mathbf{3}_S}),~~~{\cal
I}^{\varphi^{0}}_6=(\varphi^{0}(\chi\varphi)_{\mathbf{3}_S})''\xi,\\
\label{26}&&{\cal
I}^{\varphi^{0}}_7=(\varphi^{0}(\chi\varphi)_{\mathbf{3}_A})''\xi,~~~{\cal
I}^{\varphi^{0}}_8=(\varphi^{0}\chi)'\xi^2
\end{eqnarray}\vskip-0.33in
\begin{eqnarray}
\nonumber\hskip-0.3in&&{\cal
I}^{\chi^{0}}_1=(\chi^{0}\chi)(\chi\chi),~~{\cal
I}^{\chi^{0}}_2=(\chi^{0}\chi)'(\chi\chi)'',~~~{\cal
I}^{\chi^{0}}_3=(\chi^{0}\chi)''(\chi\chi)'\\
\nonumber\hskip-0.3in&&{\cal
I}^{\chi^{0}}_4=((\chi^{0}\chi)_{\mathbf{3}_S}(\chi\chi)_{\mathbf{3}_S}),~~~{\cal
I}^{\chi^{0}}_5=((\chi^{0}\chi)_{\mathbf{3}_A}(\chi\chi)_{\mathbf{3}_S}),~~~{\cal
I}^{\chi^{0}}_6=(\chi^{0}\chi)(\phi\phi),\\
\nonumber\hskip-0.3in&&{\cal
I}^{\chi^{0}}_7=(\chi^{0}\chi)'(\phi\phi)'',~~~{\cal
I}^{\chi^{0}}_8=(\chi^{0}\chi)''(\phi\phi)',~~~{\cal
I}^{\chi^{0}}_9=((\chi^{0}\chi)_{\mathbf{3}_S}(\phi\phi)_{\mathbf{3}_S})\\
\label{27}\hskip-0.3in&&{\cal
I}^{\chi^{0}}_{10}=((\chi^{0}\chi)_{\mathbf{3}_A}(\phi\phi)_{\mathbf{3}_S})
\end{eqnarray}
\begin{equation}
\label{28}\hskip-2.5in{\cal
I}^{\Delta^{0}}_1=\Delta^{0}(\phi\chi\chi),~~~{\cal
I}^{\Delta^{0}}_2=\Delta^{0}(\phi\phi\phi)
\end{equation}
\begin{equation}
\label{29}\hskip-0.75in{\cal
I}^{\rho^{0}}_1=\rho^{0}(\varphi(\phi\chi)_{\mathbf{3}_S})',~~~{\cal
I}^{\rho^{0}}_2=\rho^{0}(\varphi(\phi\chi)_{\mathbf{3}_A})',~~~{\cal
I}^{\rho^{0}}_3=\rho^{0}(\phi\chi)\xi\,.
\end{equation}
The new vacuum configuration is
obtained by imposing the vanishing of the first derivative of
$w_d+\delta w_d$ with respect to the driving fields $\varphi^{0}$,
$\chi^{0}$, $\Delta^{0}$ and $\rho^{0}$. Denoting the general flavon
field with $\Phi$, we can write the new VEV as
$\langle\Phi_i\rangle=\langle\Phi_i\rangle|_{\rm LO}+\delta
v_{\Phi_i}$. By keeping only the terms linear in the shift $\delta
v$ and neglecting the terms proportional to $\delta v/\Lambda$, the
minimization equations become:
\begin{eqnarray}
\nonumber&&(-2f_1v_{\varphi}+f_2v_{\xi})\delta v_{\varphi_3}+a_3
v_{\chi}v^2_{\varphi}/\Lambda=0\\
\nonumber&&(4f_1v_{\varphi}+f_2v_{\xi})\delta
v_{\varphi_2}+f_2v_{\varphi}\delta v_{\xi}+a_2v_{\chi}v^2_{\varphi}/\Lambda=0\\
\nonumber&&(-2f_1v_{\varphi}+f_2v_{\xi})\delta
v_{\varphi_1}+a_1v_{\chi}v^2_{\varphi}/\Lambda=0\\
\nonumber&&(M+4g_1v_{\chi})\delta v_{\chi_1}-2g_1v_{\chi}\delta
v_{\chi_2}-2g_1v_{\chi}\delta v_{\chi_3}+2g_2v_{\phi}(\delta
v_{\phi_2}-\delta v_{\phi_3})+a_4v_{\chi}v^2_{\phi}/\Lambda=0\\
\nonumber&&-2g_1v_{\chi}\delta v_{\chi_1}+4g_1v_{\chi}\delta
v_{\chi_2}+(M-2g_1v_{\chi})\delta v_{\chi_3}+2g_2v_{\phi}(\delta
v_{\phi_1}+2\delta v_{\phi_2})+a_4v_{\chi}v^2_{\phi}/\Lambda=0\\
\nonumber&&-2g_1v_{\chi}\delta v_{\chi_1}+(M-2g_1v_{\chi})\delta
v_{\chi_2}+4g_1v_{\chi}\delta v_{\chi_3}-2g_2v_{\phi}(\delta
v_{\phi_1}+2\delta v_{\phi_3})+a_4v_{\chi}v^2_{\phi}/\Lambda=0\\
\nonumber&&v_{\phi}(-\delta v_{\chi_2}+\delta
v_{\chi_3})+v_{\chi}(\delta v_{\phi_1}+\delta v_{\phi_2}+\delta
v_{\phi_3})=0\\
\label{30}&&g_4[v_{\phi}(\delta v_{\varphi_1}-\delta
v_{\varphi_3})+v_{\varphi}\delta
v_{\phi_1}]+2r_2v_{\varphi}v_{\phi}v_{\chi}/\Lambda=0\,,
\end{eqnarray}
where the parameters $a_i(i=1-4)$ are given by:
\begin{eqnarray}
\nonumber&&a_1=v_2+4v_4+2f_1(v_6+v_7)/f_2+4f^2_1v_8/f^2_2\\
\nonumber&&a_2=v_2-2v_4-2v_5+2f_1(v_6-v_7)/f_2+4f^2_1v_8/f^2_2\\
\nonumber&&a_3=v_2-2v_4+2v_5-4f_1v_6/f_2+4f^2_1v_8/f^2_2\\
\label{31}&&a_4=3(c_1+c_2+c_3)v^2_{\chi}/v^2_{\phi}-2c_6+c_7+c_8\,.
\end{eqnarray}
Eq.(\ref{30}) is linear in the shift $\delta v$ and can be
straightforwardly solved, giving:
\begin{eqnarray}
\nonumber&&\frac{\delta
v_{\varphi_1}}{v_{\varphi}}=\frac{a_1}{4f_1}\frac{v_{\chi}}{\Lambda}\\
\nonumber&&\frac{\delta
v_{\varphi_2}}{v_{\varphi}}=-\frac{f_2}{2f_1}\frac{\delta v_{\xi}}{v_{\varphi}}-\frac{a_2}{2f_1}\frac{v_{\chi}}{\Lambda}\\
\nonumber&&\frac{\delta
v_{\varphi_3}}{v_{\varphi}}=\frac{a_3}{4f_1}\frac{v_{\chi}}{\Lambda}\\
\nonumber&&\delta v_{\chi_1}=\delta v_{\chi_2}=\delta
v_{\chi_3}\equiv\delta v_{\chi}\\
\nonumber&&\frac{\delta
v_{\phi_1}}{v_{\phi}}=\Big(\frac{a_3-a_1}{4f_1}-\frac{2r_2}{g_4}\Big)\frac{v_{\chi}}{\Lambda}\\
\nonumber&&\frac{\delta
v_{\phi_2}}{v_{\phi}}=-\frac{M\delta
v_{\chi}}{4g_2v^2_{\phi}}+\Big(\frac{a_1-a_3}{8f_1}-\frac{a_4}{4g_2}+
\frac{r_2}{g_4}\Big)\frac{v_{\chi}}{\Lambda}\\
\label{32}&&\frac{\delta
v_{\phi_3}}{v_{\phi}}=\frac{M\delta
v_{\chi}}{4g_2v^2_{\phi}}+\Big(\frac{a_1-a_3}{8f_1}+\frac{a_4}{4g_2}+\frac{r_2}{g_4}\Big)\frac{v_{\chi}}{\Lambda}\,.
\end{eqnarray}
We note that the corrections to $\langle\chi\rangle$ are along the
same direction of the LO alignment, all the components of $\varphi$ acquire different corrections so that its alignment is tilted, and the shifts associated with
the components of the flavon $\phi$ are correlated with each other,
i.e., $\delta v_{\phi_1}+\delta v_{\phi_2}+\delta v_{\phi_3}=0$.
Recalling the LO relations $v_{\varphi}=-f_2v_{\xi}/(2f_1)$ and
$v^2_{\phi}=-Mv_{\chi}/(2g_2)$, the shifts $\delta v_{\xi}$ and
$\delta v_{\chi}$ can be absorbed into the redefinition of the
undetermined parameters $v_{\xi}$ and $v_{\chi}$ respectively.
Therefore the LO vacuum configuration is modified as:
\begin{eqnarray}
\nonumber&&\langle\varphi\rangle=(\delta
v_{\varphi_1},v_{\varphi}+\delta v_{\varphi_2},\delta
v_{\varphi_3}),~~~\langle\xi\rangle=v_{\xi}\\
\label{33}&&\langle\chi\rangle=(v_{\chi},v_{\chi},v_{\chi}),~~~\langle\phi\rangle=(\delta
v_{\phi_1},v_{\phi}+\delta v_{\phi_2},-v_{\phi}+\delta v_{\phi_3})\,.
\end{eqnarray}
The shifts are explicitly given in Eq.(\ref{32})
and are all of ${\cal O}(\lambda^2_c)$.

\subsection{NLO corrections to the mass matrices}

The charged lepton and neutrino mass matrices are corrected by both
the modified vacuum alignment and the subleading operators
in the superpotentials $w_{\ell}$ and $w_{\nu}$. In this section, we
present the corrections to the mass matrices and study the deviations
from TB mixing.

\subsubsection{Charged lepton}

The NLO operators contributing to the charged lepton masses can be
obtained by inserting the flavon $\chi$ in all possible ways into the LO
operators and by extracting the $A_4\times Z_4\times Z_2$ invariants;
the resulting NLO superpotential is given by:
\begin{eqnarray}
\nonumber&&\delta
w_{\ell}=\sum^2_{i=1}\frac{\tilde{y}^{(1)}_{\tau_i}}{\Lambda^2}\tau^{c}(\ell\chi\varphi)_ih_d+\frac{\tilde{y}^{(2)}_{\tau}}{\Lambda^2}\tau^{c}(\ell\chi)''\xi h_d+\sum^{5}_{i=1}\frac{\tilde{y}^{(1)}_{\mu_i}}{\Lambda^3}\mu^{c}(\ell\chi\varphi^2)_ih_d
+\sum^{2}_{i=1}\frac{\tilde{y}^{(2)}_{\mu_i}}{\Lambda^3}\mu^{c}(\ell\chi\varphi)''_i\xi
h_d\\
\nonumber&&~~~+\frac{\tilde{y}^{(3)}_{\mu}}{\Lambda^3}\mu^{c}(\ell\chi)'\xi^2h_d+\sum^{13}_{i=1}\frac{\tilde{y}^{(1)}_{e_i}}{\Lambda^4}e^{c}(\ell\chi\varphi^3)_ih_d+\sum^{5}_{i=1}\frac{\tilde{y}^{(2)}_{e_i}}{\Lambda^4}e^{c}(\ell\chi\varphi^2)''_i\xi
h_d+\sum^{2}_{i=1}\frac{\tilde{y}^{(3)}_{e_i}}{\Lambda^4}e^{c}(\ell\chi\varphi)'_i\xi^2h_d\\
\label{34}&&~~~+\frac{\tilde{y}^{(4)}_{e}}{\Lambda^4}e^{c}(\ell\chi)\xi^3h_d\,,
\end{eqnarray}
where the subscript $i$ represents different $A_4$ contractions.
The charged lepton mass matrix is obtained by adding the contributions
of this new set of operators evaluated with the insertion of the LO
VEVs of Eq.(\ref{3}) and Eq.(\ref{9}), to those of the LO
superpotential in Eq.(\ref{10}) evaluated with the NLO vacuum configuration in
Eq.(\ref{33}). After lengthy and tedious calculations, we find that
every element of charged lepton mass matrix gets corrections from both
the higher dimensional operators in $\delta w_{\ell}$ and the
shifted vacuum alignment. The off-diagonal elements become non-zero and are all suppressed by $\varepsilon$ with respect to diagonal ones.
Consequently, the corrected charged lepton mass matrix
has the following structure:
\begin{equation}
\label{35}m_{\ell}=\left(\begin{array}{ccc}m_e& \varepsilon m_e&
\varepsilon m_e\\
\varepsilon m_{\mu} & m_{\mu} & \varepsilon
m_{\mu}\\
\varepsilon m_{\tau} & \varepsilon m_{\tau}&
m_{\tau}\end{array}\right)\,,
\end{equation}
where only the order of magnitude of each non-diagonal entry is
reported. As a result, the unitary matrix $U_{\ell}$ diagonalizing $m^{\dagger}_{\ell}m_{\ell}$ is of the form:
\begin{equation}
\label{36}U_{\ell}\simeq\left(\begin{array}{ccc}1
&(V^{\ell}_{12}\varepsilon)^{*} & (V^{\ell}_{13}\varepsilon)^{*}\\
-V^{\ell}_{12}\varepsilon & 1 & (V^{\ell}_{23}\varepsilon)^{*}\\
-V^{\ell}_{13}\varepsilon & -V^{\ell}_{23}\varepsilon &1
\end{array}\right)\,,
\end{equation}
where $V^{\ell}_{ij}$ are ${\cal O}(1)$ coefficients. We note that
the charged lepton masses are corrected by terms of relative order
$\varepsilon$, thus the LO mass hierarchies are not spoiled.

\subsubsection{Neutrino}

For the heavy right-handed neutrino mass matrix $m_M$, since no
flavon field is involved in the LO Majorana mass terms of
Eq.(\ref{12}), $m_M$ does not receive corrections from the modified
vacuum alignment. Due to the strong constraint of the flavor symmetry, the
corrections to $m_M$ appear only at next to next to leading order
(NNLO), the corresponding higher dimensional operators being:
\begin{eqnarray}
\nonumber&&\frac{1}{\Lambda}\nu^{c}_1\nu^{c}_1(\phi\phi),~~~\frac{1}{\Lambda}\nu^{c}_1\nu^{c}_1(\chi\chi),~~~\frac{1}{\Lambda}\nu^{c}_1\nu^{c}_2(\chi\phi)'',
~~~\frac{1}{\Lambda}\nu^{c}_1\nu^{c}_3(\chi\phi)',~~~\frac{1}{\Lambda}\nu^{c}_2\nu^{c}_2(\phi\phi)'\\
\label{37}&&\frac{1}{\Lambda}\nu^{c}_2\nu^{c}_2(\chi\chi)',~~~\frac{1}{\Lambda}\nu^{c}_2\nu^{c}_3(\phi\phi),~~~\frac{1}{\Lambda}\nu^{c}_2\nu^{c}_3(\chi\chi),
~~~\frac{1}{\Lambda}\nu^{c}_3\nu^{c}_3(\phi\phi)'',~~~\frac{1}{\Lambda}\nu^{c}_3\nu^{c}_3(\chi\chi)''\,.
\end{eqnarray}
Given the LO vacuum alignment in Eq.(\ref{9}), we find that the 12, 13,
21, and 31 entries are still vanishing at NNLO. Taking into account the possibility of absorbing part of the corrections into the LO parameters $M$ and $M'$,
the right-handed neutrino mass matrix can be parameterized as:
\begin{equation}
\label{mmr_nlo}m_{M}=\left(\begin{array}{ccc}
M&0&0\\
0&c\varepsilon^2 M' & M'\\
0& M' & d\varepsilon^2 M'
\end{array}\right)\,,
\end{equation}
where the parameters $c$ and $d$ are of order one, their specific values are not
determined by the flavor symmetry. It is interesting to note that the mass degeneracy of the second and third right-handed neutrinos is lifted.
Then we move to consider the corrections to the Dirac neutrino mass matrix; they are
suppressed by $1/\Lambda^2$ compared to the LO and can be expressed as:
\begin{eqnarray}
\nonumber&&\delta
w_{\nu}=\frac{y_{\nu_1}}{\Lambda}\nu^{c}_1(\ell\delta\phi)h_u+\frac{y_{\nu_2}}{\Lambda}\nu^{c}_2(\ell\delta\chi)''h_u+\frac{y_{\nu_3}}{\Lambda}\nu^{c}_3(\ell\delta\chi)'h_u
+\frac{\tilde{y}_{\nu_4}}{\Lambda^2}\nu^{c}_1(\ell(\chi\phi)_{\mathbf{3}_S})h_u\\
\nonumber&&~~~~~+\frac{\tilde{y}_{\nu_5}}{\Lambda^2}\nu^{c}_1(\ell(\chi\phi)_{\mathbf{3}_A})h_u+\frac{\tilde{y}_{\nu_6}}{\Lambda^2}\nu^{c}_2(\ell\chi\chi)''h_u+\frac{\tilde{y}_{\nu_7}}{\Lambda^2}\nu^{c}_2(\ell\phi\phi)''h_u
+\frac{\tilde{y}_{\nu_8}}{\Lambda^2}\nu^{c}_3(\ell\chi\chi)'h_u\\
\label{38}&&~~~~~+\frac{\tilde{y}_{\nu_9}}{\Lambda^2}\nu^{c}_3(\ell\phi\phi)'h_u\,,
\end{eqnarray}
where $\delta\phi$ and $\delta\chi$ denote the shifted vacuum of the
flavons $\phi$ and $\chi$, respectively. Since the shift
$\delta\chi$ turns out to be proportional to the LO VEV and the
symmetric triplet
$(\phi\phi)_{\mathbf{3}_S}=2(\phi^2_1-\phi_2\phi_3,\phi^2_3-\phi_1\phi_2,\phi^2_2-\phi_1\phi_3)$
has a VEV in the same direction as $\langle\chi\rangle$, the contributions of the
terms proportional to $\delta\chi$, $\tilde{y}_{\nu_7}$ and
$\tilde{y}_{\nu_9}$ can be absorbed into the redefinition of the
parameters $y_{\nu_2}$ and $y_{\nu_3}$. Moreover, the fourth term
can be absorbed by a redefinition of $y_{\nu_1}$ whereas the operators
with coefficients $\tilde{y}_{\nu_6}$ and $\tilde{y}_{\nu_8}$ give a
vanishing contribution. Therefore the relevant correction to the Dirac mass matrix comes from the above terms proportional to $y_{\nu_1}$ and
$\tilde{y}_{\nu_5}$, giving:
\begin{equation}
\label{39}\delta
m_{D}=\left(\begin{array}{ccc}\tilde{y}&-\tilde{y}&0\\
0&0&0\\
0&0&0
\end{array}\right)\frac{v_{\chi}v_{\phi}}{\Lambda^2}v_u\,,
\end{equation}
where $\tilde{y}=y_{\nu_1}\Lambda\delta
v_{\phi_1}/(v_{\chi}v_{\phi})-2\tilde{y}_{\nu_5}$. Then the NLO
correction to the light neutrino mass matrix is given by:
\begin{eqnarray}
\nonumber&&\delta m_{\nu}=-\delta
m^{T}_Dm^{-1}_{M}m_D-m^{T}_Dm^{-1}_{M}\delta m_D=\\
\label{40}&&=\tilde{y}y_{\nu_1}\left(\begin{array}{ccc}0&1&-1\\
1&-2&1\\
-1&1&0
\end{array}\right)\frac{v^2_{\phi}v_{\chi}}{\Lambda^2M}\frac{v^2_u}{\Lambda}\,.
\end{eqnarray}
Diagonalizing the modified light neutrino mass matrix, we find that the first light neutrino is still massless and a non-zero mass only arises
at NNLO. Combining the NLO corrections from the charged lepton and
neutrino sectors, the parameters of the lepton mixing matrix are
modified as:
\begin{eqnarray}
\nonumber&&\sin\theta_{13}=\Big|\frac{\tilde{y}}{\sqrt{2}\,y_{\nu_1}}\frac{v_{\chi}}{\Lambda}+\frac{1}{\sqrt{2}}(V^{\ell}_{12}-V^{\ell}_{13})\varepsilon\Big|\\  
\nonumber&&\sin^2\theta_{12}=\frac{1}{3}-\frac{1}{3}[(V^{\ell}_{12}+V^{\ell}_{13})\varepsilon+h.c.]\\
\label{41}&&\sin^2\theta_{23}=\frac{1}{2}+\frac{1}{4}[\frac{\tilde{y}}{y_{\nu_1}}\frac{v_{\chi}}{\Lambda}+2V^{\ell}_{23}\varepsilon+h.c.] \,. 
\end{eqnarray}
All the three mixing angles receive corrections of order
$\lambda^2_c$, the deviation of solar angle from its TB value is controlled by the flavor mixing
in the charged lepton sector and the reactor
angle $\theta_{13}$ is expected to be of order $\lambda^2_c$.
Recently the T2K collaboration reported a relative large value for
$\theta_{13}$ \cite{Abe:2011sj}. This result, combined with the world neutrino data, gives the
$3\sigma$ ranges of $\sin^2\theta_{13}$ as $[0.001,0.044]$ and $[0.005,0.050]$, for the so-called "old" and "new" reactor neutrino
flux, respectively \cite{Fogli:2011qn}. Then values of $\theta_{13}\sim {\cal O}(\lambda^2_c)$ lie within these ranges and our model
cannot be ruled out, as is shown in Fig. \ref{fig:mixing_angles}. Precise measurement of $\theta_{13}$
is an important test of this model: if a large $\theta_{13}$ close
to the present upper bound is confirmed by future data, then our construction would
be ruled out. The same remark applies to a large class of recent
discrete flavor symmetry models.

\section{Phenomenological implications}
In this section, we study the predictions for leptogenesis and lepton flavor violation both analytically and numerically.

\subsection{Leptogenesis}
It is interesting to estimate the order of magnitude of the right-handed neutrino masses.
Recalling the light neutrino masses given in Eq.(\ref{19}) and taking the couplings $y_{\nu_1}$, $y_{\nu_2}$ and $y_{\nu_3}$ to be of ${\cal O}(1)$ and the VEVs $v_{\chi}/\Lambda$ and $v_{\phi}/\Lambda$ of  ${\cal O}(\lambda^2_c)$, we obtain:
\begin{equation}
M\sim M'\sim10^{12\div13}{\rm GeV}\,.
\end{equation}
It has been established that flavor effects may play an important role in leptogenesis
\cite{flavor_leptogenesis1,flavor_leptogenesis2,flavor_leptogenesis3,flavor_leptogenesis4,flavor_leptogenesis5}.
If the right-handed neutrino masses are larger than $(1+\tan^2\beta)\times10^{12}$ GeV, with $\tan\beta\equiv v_u/v_d$ being the ratio of the vacuum expectation values
of the two Higgs doublets in the minimal supersymmetric standard model (MSSM), the three flavors $e$, $\mu$ and $\tau$ are indistinguishable and the so-called
"one-flavor" approximation can be safely used. For $(1+\tan^2\beta)\times10^{9}{\rm GeV}\ll M(M')\ll(1+\tan^2\beta)\times10^{12}{\rm GeV}$, only the $\tau$ Yukawa coupling is
in equilibrium and should be treated separately in the Boltzmann equations, while the $e$ and $\mu$ flavors are still indistinguishable. On the other hand, for $(1+\tan^2\beta)\times10^{5}{\rm GeV}\ll M(M')\ll(1+\tan^2\beta)\times10^{9}{\rm GeV}$, the charged $\mu$ and $\tau$ Yukawa couplings are in thermal equilibrium and all flavors should be treated separately. For natural values of the parameters, e.g., $\tan\beta<30$ and the neutrino Yukawa coupling $y_{\nu_i}$ of ${\cal O}(1)$, our model lies in the flavored regime where the $\tau$ flavor should be considered separately from the others. 

The implication of the $A_4$ group for leptogenesis has been discussed extensively
\cite{Branco:2009by,Bertuzzo:2009im,Hagedorn:2009jy,Jenkins:2008rb,leptogenesis_A4}. In general, the leptonic CP asymmetries are predicted to be vanishing at LO,
since the combination $Y^{\nu}Y^{\nu\dagger}$, which is relevant for leptogenesis, is proportional to the unit matrix, where $Y^{\nu}=m_D/v_{u}$ is the neutrino Yukawa
coupling matrix. Thus subleading operators, suppressed by additional powers of the cutoff $\Lambda$, are required to account for leptogenesis.
It is well known that the leptonic CP asymmetry parameters $\epsilon^{\alpha}_i$ for the $i-$th heavy right-handed (s)neutrino $\nu^{c}_i$ ($\tilde{\nu}^{c}_i$)
decaying into $\alpha-$lepton ($\alpha=e,\mu,\tau$), provided the heavy neutrino masses are far from being almost degenerate, are given by \cite{asymmetry}:
\begin{eqnarray}
\epsilon^{\alpha}_i=\frac{1}{8\pi}\sum_{j\neq i}\frac{{\rm Im}[(\hat{Y}^{\nu}\hat{Y}^{\nu\dagger})_{ij}
\hat{Y}^{\nu}_{i\alpha}\hat{Y}^{\nu*}_{j\alpha}]}{(\hat{Y}^{\nu}\hat{Y}^{\nu\dagger})_{ii}}g\left(\frac{M^2_j}{M^2_i}\right)\,,
\end{eqnarray}
with the loop function $g$ expressed as:
\begin{equation}
g(x)=\sqrt{x}\left[\frac{2}{1-x}-\ln\left(\frac{1+x}{x}\right)\right]\,,
\end{equation}
where the {\it hat} denotes the basis in which the mass matrices $m_M$ and $m_{\ell}$ are diagonal with real and non-negative entries.
On the other hand, for an almost degenerate heavy neutrino mass spectrum, leptogenesis can be naturally implemented through the so-called resonant
leptogenesis mechanism \cite{resonant:leptogenesis}. In this case, the CP asymmetry generated by the decay of the $i-$th heavy right-handed (s)neutrino
$\nu^{c}_i$ ($\tilde{\nu}^{c}_i$) into a lepton flavor $\alpha$ is given by \cite{resonant:leptogenesis,flavor_leptogenesis5}:
\begin{equation}
\label{eq:rs}\epsilon^{\alpha}_i=-\frac{1}{8\pi}\sum_{j\neq i}\frac{M_iM_j\Delta M^2_{ij}}{(\Delta M^2_{ij})^2+M^2_i\Gamma^2_j}
\frac{{\rm Im}[(\hat{Y}^{\nu}\hat{Y}^{\nu\dagger})_{ij}\hat{Y}^{\nu}_{i\alpha}\hat{Y}^{\nu*}_{j\alpha}]}{(\hat{Y}^{\nu}\hat{Y}^{\nu\dagger})_{ii}}\,,
\end{equation}
where $\Delta M^2_{ij}=M^2_j-M^2_i$ and $\Gamma_j=(\hat{Y}^{\nu}\hat{Y}^{\nu\dagger})_{jj}M_{j}/(8\pi)$ is the decay width of the $j-$th right-handed neutrino. In our model, the resonant leptogenesis mechanism is only applicable to the second and third heavy neutrinos, see Eq.(\ref{14}).

Since we work in the hatted basis, we have to consider the diagonalization of right-handed neutrino mass matrix $m_M$ of Eq.(\ref{mmr_nlo});
it is diagonalized as $\widetilde{U}^{T}_Rm_M\widetilde{U}_R={\rm diag}(M_1, M_2, M_3)$, with the mass eigenvalues
\begin{eqnarray}
\label{eq:mass}M_1=M,~~~M_2\simeq\left[1+\frac{1}{2}(c+d)\varepsilon^2\right]M',~~~M_3\simeq\left[1-\frac{1}{2}(c+d)\varepsilon^2\right]M'\,,
\end{eqnarray}
where we take the parameters $c$ and $d$ to be real for simplicity, the complex case follows analogously.
The matrix $\widetilde{U}_R$ can be written as:
\begin{equation}
\widetilde{U}_R\simeq\frac{1}{\sqrt{2}}\left(\begin{array}{ccc}
\sqrt{2} &0 &0\\
0& 1+\frac{1}{4}(c-d)\varepsilon^2 & -i[1-\frac{1}{4}(c-d)\varepsilon^2]\\
9& 1-\frac{1}{4}(c-d)\varepsilon^2& i[1+\frac{1}{4}(c-d)\varepsilon^2]
\end{array}\right)\,;
\end{equation}
consequently, in the hatted basis, the neutrino Yukawa coupling matrix is:
\begin{eqnarray}
\label{eq:ynu}\hat{Y}^{\nu}=\frac{1}{v_{u}}\widetilde{U}^{T}_R(m_D+\delta m_D)U_{\ell}=\frac{1}{\sqrt{2}}\left(\begin{array}{ccc}
0&-\sqrt{2}\,x & \sqrt{2}\,x\\
y+z & y+z & y+z\\
i(-y+z) & i(-y+z) & i(-y+z)
\end{array}\right)+{\cal O}(\varepsilon^2)\,,
\end{eqnarray}
where, for simplicity, we denoted $x\equiv y_{\nu_1}v_{\phi}/\Lambda$, $y\equiv y_{\nu_2}v_{\chi}/\Lambda$ and $z\equiv y_{\nu_3}v_{\chi}/\Lambda$.
The leptogenesis is associated with both $\hat{Y}^{\nu}$ and the combination $\hat{Y}^{\nu}\hat{Y}^{\nu\dagger}$, which reads:
\begin{eqnarray}
\label{eq:ynucomb}\hat{Y}^{\nu}\hat{Y}^{\nu\dagger}=\left(\begin{array}{ccc}
|x|^2+|w|^2+|x+w|^2 &0 &0 \\
0&3/2|y+z|^2 & 3i/2(y+z)(y^{*}-z^{*})\\
0& -3i/2(y-z)(y^{*}+z^{*}) & 3/2|y-z|^2
\end{array}\right)+{\cal O}(\varepsilon^4)\,,
\end{eqnarray}
where $w=\tilde{y}v_{\chi}v_{\phi}/\Lambda^2$ comes from the NLO correction $\delta m_D$.
It is remarkable that $(\hat{Y}^{\nu}\hat{Y}^{\nu\dagger})_{12}=(\hat{Y}^{\nu}\hat{Y}^{\nu\dagger})_{13}=(\hat{Y}^{\nu}\hat{Y}^{\nu\dagger})_{21}=
(\hat{Y}^{\nu}\hat{Y}^{\nu\dagger})_{31}\simeq0$ even if the NLO corrections are taken into account. As a result, we have:
\begin{equation}
\epsilon^{\alpha}_1\simeq0\,.
\end{equation}
This implies that the heavy neutrino $\nu^{c}_1$ decouples and the CP violating lepton asymmetry is produced in the out of equilibrium decays
of the heavy neutrinos $\nu^{c}_2$ and $\nu^{c}_3$.
Combining the expression in Eq.(\ref{eq:rs}) with Eqs.(\ref{eq:mass},\ref{eq:ynu},\ref{eq:ynucomb}), the flavor dependent CP asymmetry parameters
are as follows:
\begin{eqnarray}
\nonumber&&\epsilon^{e}_2\simeq\epsilon^{\mu}_2\simeq\epsilon^{\tau}_2\simeq\frac{1}{2\pi}\frac{(c+d)\varepsilon^2}{4(c+d)^2\varepsilon^4+\frac{9}{256\pi^2}|y-z|^4}\frac{|y|^2-|z|^2}{|y+z|^2}\,{\rm Im}(yz^{*})\\
\label{eq:cp}&&\epsilon^{e}_3\simeq\epsilon^{\mu}_3\simeq\epsilon^{\tau}_3\simeq\frac{1}{2\pi}\frac{(c+d)\varepsilon^2}{4(c+d)^2\varepsilon^4+
\frac{9}{256\pi^2}|y+z|^4}\frac{|y|^2-|z|^2}{|y-z|^2}\,{\rm Im}(yz^{*})\,.
\end{eqnarray}
It is interesting to note that all the parameters are proportional to the combination $(|y|^2-|z|^2){\rm Im}(yz^{*})$
so that the $\epsilon^{\alpha}_i$'s would be vanishing in the limit of $|y|=|z|$ or ${\rm arg}(y)={\rm arg}(z)$.
Besides the above $\epsilon^{\alpha}_i$'s, the baryon asymmetry depends on the so-called wash-out mass parameters $\widetilde{m}^{\alpha}_i$
associated with each lepton asymmetry:
\begin{equation}
\widetilde{m}^{\alpha}_i=\frac{|\hat{Y}^{\nu}_{i\alpha}|^2v^2_u}{M_i}\,.
\end{equation}
Then we have:
\begin{eqnarray}
\nonumber&&\widetilde{m}^{e}_2\simeq\widetilde{m}^{\mu}_2\simeq\widetilde{m}^{\tau}_2\simeq\frac{|y+z|^2v^2_u}{[2+(c+d)\varepsilon^2]M'}\,,\\
&&\widetilde{m}^{e}_3\simeq\widetilde{m}^{\mu}_3\simeq\widetilde{m}^{\tau}_3\simeq\frac{|y-z|^2v^2_u}{[2-(c+d)\varepsilon^2]M'}\,.
\end{eqnarray}
Once the values of the CP parameters $\epsilon^{\alpha}_i$ is fixed, the final value of baryon asymmetry $Y_B$ is governed by a set of flavor-dependent Boltzmann equations including the (inverse) decay and scattering process as well as the nonperturbative sphaleron interaction \cite{Davidson:2008bu}. Here we will use simple analytical formulae to estimate baryon asymmetry \cite{flavor_leptogenesis3,flavor_leptogenesis5,Blanchet:2006dq}:
\begin{equation}
Y_{B}\simeq\sum^{3}_{i=2}Y_{B_i}\simeq-\frac{10}{31g_{*}}\sum^{3}_{i=2}\left[\epsilon^{e+\mu}_i\eta\Big(\frac{417}{589}\widetilde{m}^{e+\mu}\Big)+
\epsilon^{\tau}_i\eta\Big(\frac{390}{589}\widetilde{m}^{\tau}\Big)\right]\,,
\end{equation}
where $\epsilon^{e+\mu}_i=\epsilon^{e}_i+\epsilon^{\mu}_i$, $\widetilde{m}^{e+\mu}=\widetilde{m}^{e}_2+\widetilde{m}^{\mu}_2+\widetilde{m}^{e}_3+\widetilde{m}^{\mu}_3$ and $\widetilde{m}^{\tau}=\widetilde{m}^{\tau}_2+\widetilde{m}^{\tau}_3$, $g_{*}=228.75$ is the effective number of degrees of freedom in the MSSM. We note that the wash-out mass parameters are added up since the asymmetry generated
in $N_3$ decays can be washed out by $N_2$ interactions and vice versa \cite{Blanchet:2006dq}. The wash-out factor $\eta(\widetilde{m}^{\alpha})$ accounts for the washing out effect of the total baryon asymmetry due to the inverse decays and $\Delta L=1$ scattering, its explicit expression depends on the magnitude of the various $\widetilde{m}^{\alpha}$. If all the flavors are in the strong wash-out regime, or some flavors are strongly washed out and others are either weakly or mildly washed out, it is given by \cite{flavor_leptogenesis3,flavor_leptogenesis5}:
\begin{equation}
\eta(\widetilde{m}^{\alpha})=\left[\frac{8.25\times10^{-3}{\rm eV}}{\widetilde{m}^{\alpha}}+\Big(\frac{\widetilde{m}^{\alpha}}{0.2\times10^{-3}{\rm eV}}\Big)^{1.16}\right]^{-1}\,.
\end{equation}
While if all the flavors are in the weak wash-out regime, $\eta(\widetilde{m}^{\alpha})$ is well approximated by \cite{flavor_leptogenesis3}
\begin{equation}
\eta(\widetilde{m}^{\alpha})=1.5\left(\frac{\widetilde{m}}{3.3\times10^{-3}\mathrm{eV}}\right)\left(\frac{\widetilde{m}^{\alpha}}{3.3\times10^{-3}\mathrm{eV}}\right)\,,
\end{equation}
where $\widetilde{m}=\sum_{\alpha}\widetilde{m}^{\alpha}$.

\subsection{Lepton flavor violation}
We perform the analysis within the framework of the minimal supergravity (mSUGRA) scenario, which provides flavor universal
boundary conditions at the scale of grand unification $M_G\simeq2\times10^{16}$GeV. It assumes that the slepton mass matrices are diagonal and universal in flavor and the trilinear couplings are proportional to the Yukawa couplings at the scale $M_G$. The branching ratio of the lepton flavor violation (LFV) radiative decay $\ell_i\rightarrow\ell_j+\gamma$ is
approximately given by \cite{Petcov:2003zb,Petcov:2005jh}:
\begin{equation}
Br(\ell_i\rightarrow\ell_j+\gamma)\simeq Br(\ell_i\rightarrow\ell_j+\nu_i+\bar{\nu}_j)\frac{\alpha^3_{em}}{G^2_Fm^8_S}
\Big(\frac{3m^2_0+A^2_0}{8\pi^2}\Big)^2\Big|(\hat{Y}^{\nu\dagger}{\mathbf{L}}\hat{Y}^{\nu})_{ij}\Big|^2\tan^2\beta\,,
\end{equation}
where $G_F$ is the Fermi constant and $\alpha_{em}$ is the fine structure constant, $m_0$ is the common scalar mass, $A_0$ is
 the common trilinear parameter and the factor $\mathbf{L}$ is given by:
\begin{equation}
\mathbf{L}_{ij}=\ln\left(\frac{M_G}{M_i}\right)\delta_{ij}\,.
\end{equation}
The parameter $m_S$ is the character mass scale of the SUSY particle and an excellent approximation to the exact result is given by \cite{Petcov:2003zb}:
\begin{equation}
m^8_S\simeq0.5m^2_0m^2_{1/2}(m^2_0+0.6m^2_{1/2})^2\,,
\end{equation}
where $m_{1/2}$ is the universal gaugino mass. Recalling the neutrino Yukawa coupling matrix $\hat{Y}^{\nu}$ given in Eq.(\ref{eq:ynu}), we straightforwardly have:
\begin{eqnarray}
\nonumber&&(\hat{Y}^{\nu\dagger}\mathbf{L}\hat{Y}^{\nu})_{12}=(\hat{Y}^{\nu\dagger}\mathbf{L}\hat{Y}^{\nu})_{21}=(|y|^2+|z|^2)\ln\left(\frac{M_G}{M'}\right)+{\cal O}(\varepsilon^3)\\
\nonumber&&(\hat{Y}^{\nu\dagger}\mathbf{L}\hat{Y}^{\nu})_{13}=(\hat{Y}^{\nu\dagger}\mathbf{L}\hat{Y}^{\nu})_{31}=(|y|^2+|z|^2)\ln\left(\frac{M_G}{M'}\right)+{\cal O}(\varepsilon^3)\\
\label{eq:YLY}&&(\hat{Y}^{\nu\dagger}\mathbf{L}\hat{Y}^{\nu})_{23}=(\hat{Y}^{\nu\dagger}\mathbf{L}\hat{Y}^{\nu})_{32}=-|x|^2\ln\left(\frac{M_G}{M}\right)+(|y|^2+|z|^2)\ln\left(\frac{M_G}{M'}\right)
+{\cal O}(\varepsilon^3)\,.
\end{eqnarray}
It is remarkable that the relation $(\hat{Y}^{\nu\dagger}\mathbf{L}\hat{Y}^{\nu})_{21}=(\hat{Y}^{\nu\dagger}\mathbf{L}\hat{Y}^{\nu})_{31}$ holds at LO,
which is related to the $\mu-\tau$ symmetry of the light neutrino mass matrix. The same result has been obtained in previous $A_4$ and $S_4$ models
\cite{Ding:2009gh,Hagedorn:2009df}. As a consequence, the LFV branching ratios are as follows:
\begin{equation}
\label{eq:LFV_relation}Br(\tau\rightarrow e\gamma)\simeq Br(\tau\rightarrow e\nu_{\tau}\bar{\nu}_e)Br(\mu\rightarrow e\gamma)\simeq0.18Br(\mu\rightarrow e\gamma)\,.
\end{equation}
Given the latest experimental bound $Br(\mu\rightarrow e\gamma)<2.4\times10^{-12}$ \cite{meg}, the rate of $\tau\rightarrow e\gamma$ should be much below the present and future sensitivities \cite{pdg}. Therefore, if the future experiments at SuperB factory
will find $Br(\tau\rightarrow e\gamma)<10^{-9}$, we would not constrain any further the present $A_4$ model.
Otherwise the observation of $\tau\rightarrow e\gamma$ with branching ratio $\geq10^{-9}$, combined with upper limit on
$Br(\mu\rightarrow e\gamma)$, would rule out this model. Notice that, due to the $-|x|^2\ln(M_G/M)$ term in the third equation of Eq.(\ref{eq:YLY}) and to the fact that the scale $M$ is generally smaller than the GUT scale $M_G$, the branching ratio $Br(\tau\rightarrow\mu\gamma)$ is not linearly related to $Br(\mu\rightarrow e\gamma)$ and should be smaller than $Br(\tau\rightarrow e\gamma)$.

Tripleton decays $\ell_i\rightarrow3\ell_j$ and $\mu-e$ conversion in nuclei are generally related to the previous LFV radiative decays.
In the mSUGRA scenario, the LFV processes are dominated by the contributions coming from the
$\gamma-$penguin diagrams. As a consequence, the branching ratio for trilepton decays $\ell_i\rightarrow3\ell_j$ is approximately given by \cite{Hisano:1995cp}:
\begin{equation}
Br(\ell_i\rightarrow3\ell_j)\simeq\frac{\alpha_{em}}{3\pi}[\;\ln(\frac{m^2_{\ell_i}}{m^2_{\ell_j}})-\frac{11}{4}]Br(\ell_i\rightarrow\ell_j\gamma)\,.
\end{equation}
Concretely, we have $Br(\mu\rightarrow3e)\simeq0.006Br(\mu\rightarrow e\gamma)$, $Br(\tau\rightarrow3e)\simeq0.01Br(\tau\rightarrow e\gamma)$ and $Br(\tau\rightarrow3\mu)\simeq0.002Br(\tau\rightarrow\mu\gamma)$. For $\mu-e$ conversion in nuclei,
the $\gamma-$penguin dominance implies \cite{Hisano:1995cp}:
\begin{equation}
CR(\mu N\rightarrow eN)=\frac{\Gamma(\mu N\rightarrow eN)}{\Gamma_{cap}}\simeq\frac{\alpha^4_{em}m^5_{\mu}G^2_F}{12\pi^3\Gamma_{cap}}ZZ^4_{eff}|F(q^2)|^2
Br(\mu\rightarrow e\gamma)\,,
\end{equation}
where $\Gamma_{cap}$ is the experimentally measured total muon capture rate, $Z$ is the proton number in the nucleus,
$Z_{eff}$ is the effective atomic charge obtained by averaging the muon wavefunction over the
nuclear density, and $F(q^2)$ denotes the nuclear form factor at momentum transfer $q$. For $^{48}_{22}$Ti, we have $Z_{eff}=17.6$, $F(q^2\simeq-m^2_{\mu})\simeq0.54$ and $\Gamma_{cap}=1.70422\times10^{-18}$ GeV \cite{prime}. In the case of $^{27}_{13}$Al, one finds $Z_{eff}=11.5$, $F(q^2\simeq-m^2_{\mu})\simeq0.64$ and $\Gamma_{cap}=4.64079\times10^{-19}$ GeV \cite{mu2e}. As a result, the $\mu-e$ conversion rates in $^{48}_{22}$Ti and $^{27}_{13}$Al are given by
\begin{equation}
CR(\mu\;^{48}_{22}{\rm Ti}\rightarrow e\;^{48}_{22}{\rm Ti})\simeq0.0049Br(\mu\rightarrow e\gamma),~~CR(\mu\;^{27}_{13}{\rm Al}\rightarrow
e\;^{27}_{13}{\rm Al})\simeq0.0027Br(\mu\rightarrow e\gamma)\,.
\end{equation}
The sensitivity of future $\mu-e$ conversion experiments in $^{48}_{22}$Ti and $^{27}_{13}$Al will be improved drastically
to $10^{-18}$ and $10^{-16}$ respectively and this corresponds to the upper bounds on $Br(\mu\rightarrow e\gamma)$ of $2.04\times10^{-16}$ and $3.7\times10^{-14}$,
smaller than the prospected sensitivity $Br(\mu\rightarrow e\gamma)<10^{-13}$ in the MEG experiment \cite{meg}.
Therefore the $\mu-e$ conversion experiments can further constrain the model if $\mu\rightarrow e\gamma$ is not observed by MEG.

\subsection{Numerical results}
\begin{figure}[hptb]
\begin{center}
\begin{tabular}{cc}
\includegraphics[scale=.36]{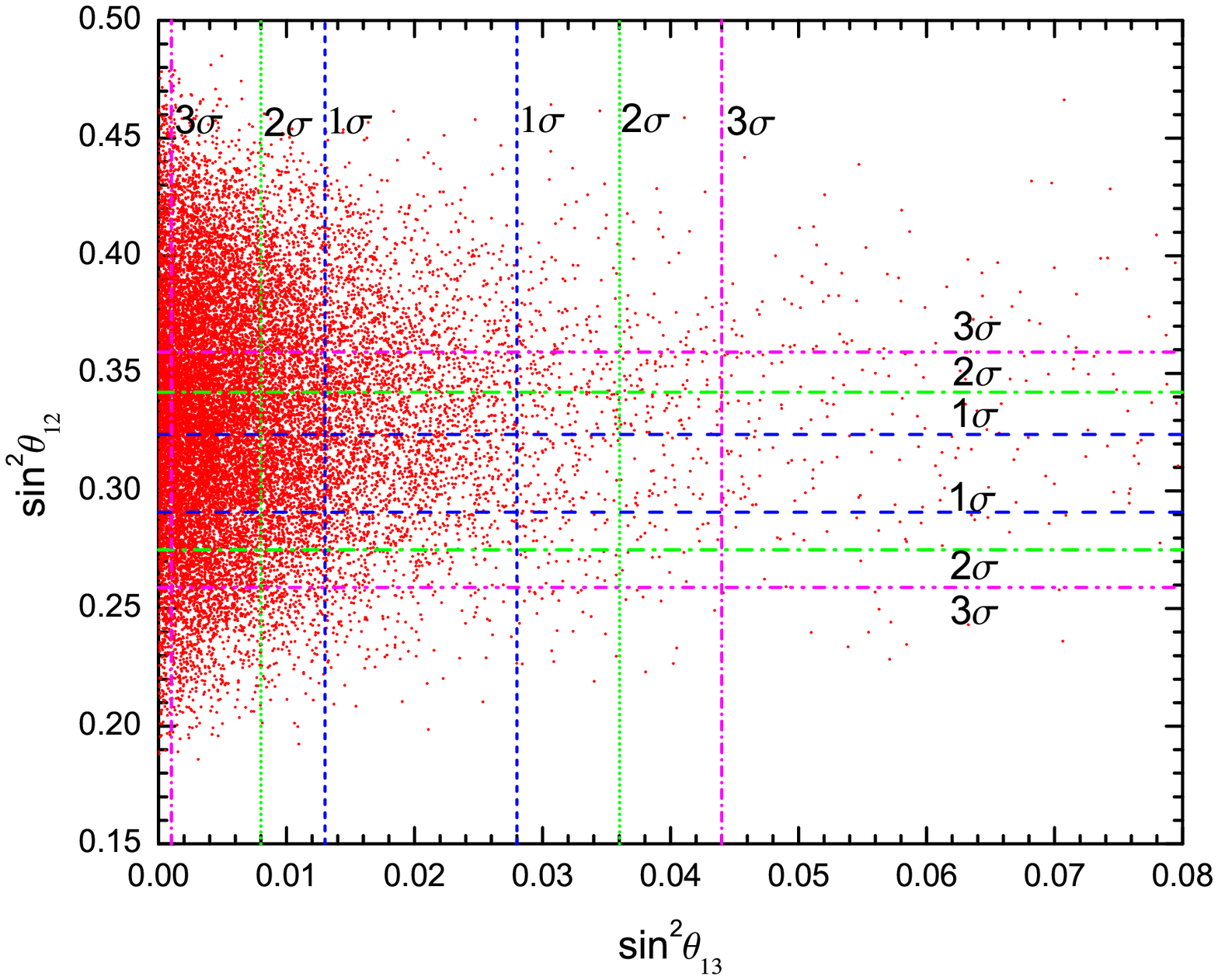}&\includegraphics[scale=.36]{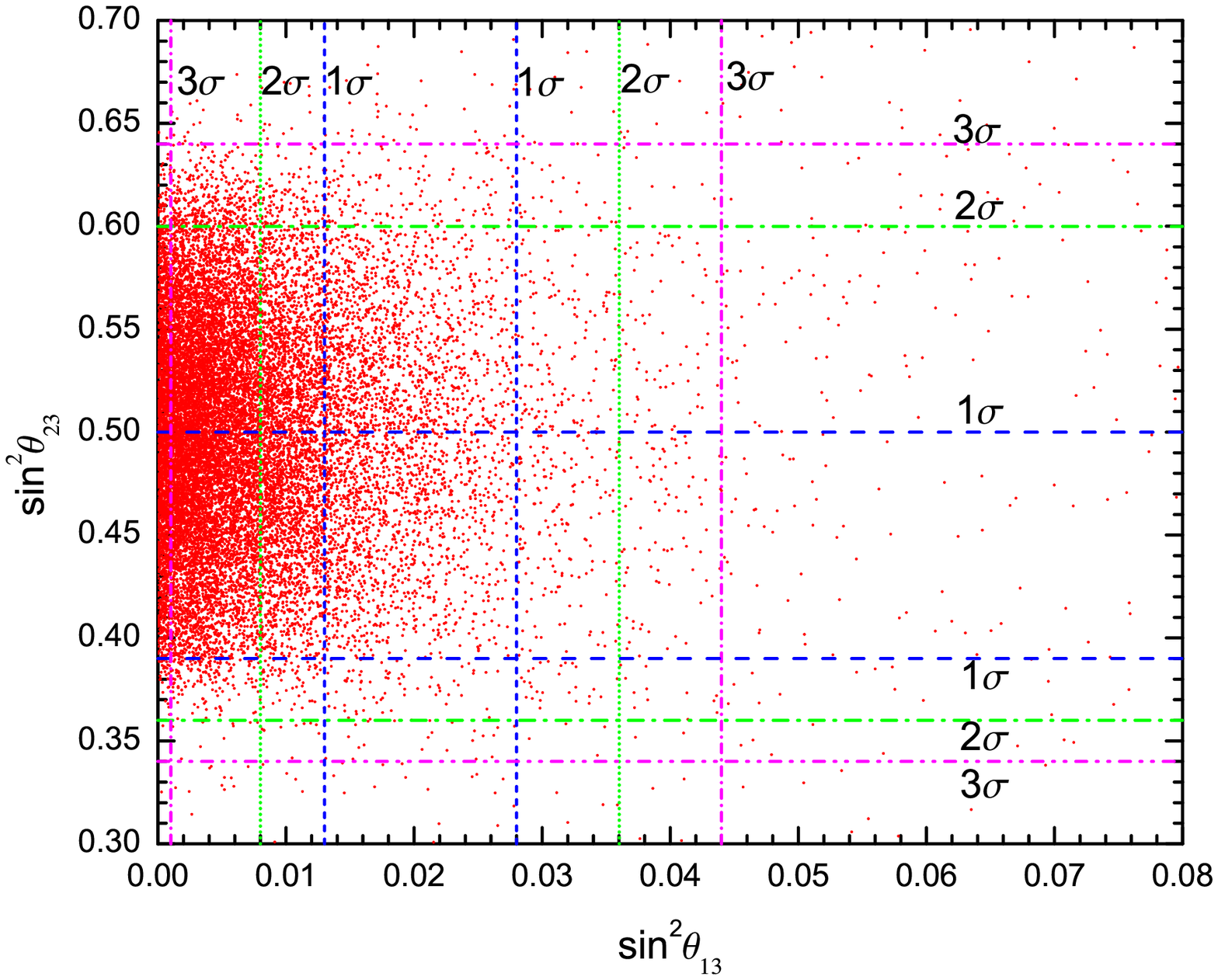}
\end{tabular}
\caption{\label{fig:mixing_angles} \it The scatter plot of $\sin^2\theta_{12}$ and $\sin^2\theta_{23}$ against $\sin^2\theta_{13}$. The
$1\sigma$, $2\sigma$ and $3\sigma$ bounds on the mixing angles are taken from Ref. \cite{Fogli:2011qn} with old reactor neutrino fluxes.}
\end{center}
\end{figure}

\begin{figure}[hptb]
\begin{center}
\includegraphics[scale=0.40]{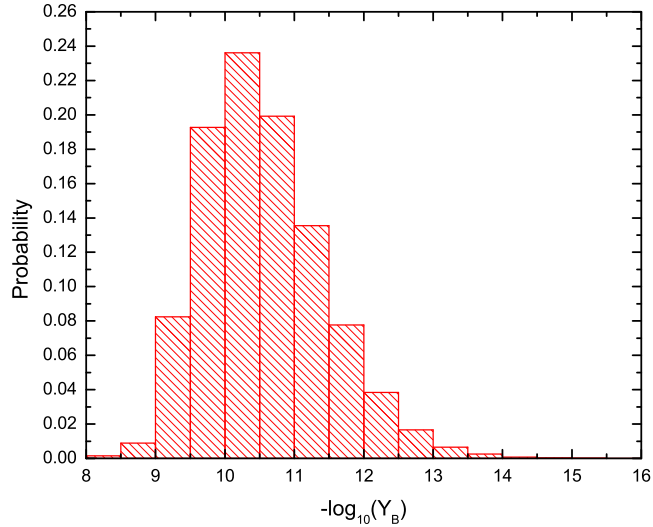}
\caption{\label{fig:leptogensis}\it The distribution of the final baryon asymmetry $Y_B$ in the limit $y=z$.}
\end{center}
\end{figure}

In order to see more clearly the phenomenological implications of the model and to check the LO analytical results, we performed a numerical analysis. Since the parameters $x$, $y$ and $z$ of Eq.(\ref{eq:ynu}) are expected to be of order $\lambda^2_c$, they are treated as random
complex numbers with absolute value between 0.01 and 0.1; the absolute value of the NLO parameter $w$ varies in the range of
$[0.01\lambda^2_c, 0.1\lambda^2_c]$ and the corresponding phase between 0 and $2\pi$. The parameters $V^{\ell}_{12}$, $V^{\ell}_{13}$ and $V^{\ell}_{23}$ in the lepton mixing matrix $U_{\ell}$ of Eq.(\ref{36}) and the parameters $c$ and $d$ in the NNLO Majorana neutrino mass matrix $m_M$ are taken to be complex numbers with absolute value in the interval $[1/3,3]$, the heavy neutrino mass parameters $M$ and $M'$ are
allowed to vary from $10^{11}$ GeV to $10^{14}$ GeV and the expansion parameter $\varepsilon$ is set to the indicative value 0.04. 
The correlations of $\sin^2\theta_{12}$ and $\sin^2\theta_{23}$ with respect to $\sin^2\theta_{13}$ are shown in Fig.\ref{fig:mixing_angles}.
In these plots, we require that the corresponding $\Delta m^2_{sol}$ and $\Delta m^2_{atm}$ are within the $3\sigma$ ranges taken from Ref. \cite{Fogli:2011qn}. We clearly see that most of the points fall in the region where $\sin^2\theta_{12}$ and $\sin^2\theta_{23}$ are in the
$3\sigma$ interval; the values of $\theta_{13}$ are consistent with the global neutrino data analysis including the T2K results. However, if $\theta_{13}$ is measured to be near the present upper bound in future experiments, this model would be almost ruled out.
Next we move to discuss the numerical results for leptogenesis parameters and LFV branching fractions. For definiteness,
we shall present our results only for the mSUGRA point SPS3 \cite{Allanach:2002nj}. The SPS3 point is in the co-annihilation region for the SUSY dark matter and the values
of the universal soft SUSY breaking parameters are as follows:
\begin{equation}
m_0=90\, {\rm GeV},~~m_{1/2}=400 \,{\rm GeV}, ~~A_0=0\, {\rm GeV},~~\tan\beta=10\,.
\end{equation}
We note that only the parameter $\tan\beta$ is relevant for leptogenesis. Our detailed numerical analysis shows that the observed baryon asymmetry can be obtained by
requiring a moderate cancellation between $y$ and $z$ in the common factor $(|y|^2-|z|^2){\rm Im}(yz^{*})$ of the leptonic CP asymmetries given by Eq.(\ref{eq:cp}).
In the extreme case of $y=z$, the distribution of the predicted $Y_B$ is plotted in Fig. \ref{fig:leptogensis}. The plot has been done by taking into account higher order contributions not explicitly shown in Eq.(\ref{eq:cp}).
It is interesting that the resulting
baryon asymmetry $Y_B$ is rather small and a sizable
part of points falls into the region where $Y_B$ is in the phenomenologically allowed interval of $[10^{-11},10^{-10}]$, while the predictions for mixing angles and
LFV branching ratios are essentially the same as the general $y\neq z$ case. We note that the scenario $y=z$ could
be realized by extending the $A_4$ flavor symmetry to $S_4$ and unifying the second and third right-handed neutrinos into an $S_4$ doublet. Considering the fact that
$A_4$ is a normal subgroup of $S_4$ and the doublet representation of $S_4$ decomposes into $1'$ and $1''$ representations of $A_4$, the resulting model would be very
similar to the present one. 

\begin{figure}[hptb]
\begin{center}
\begin{tabular}{cc}
\includegraphics[scale=.30]{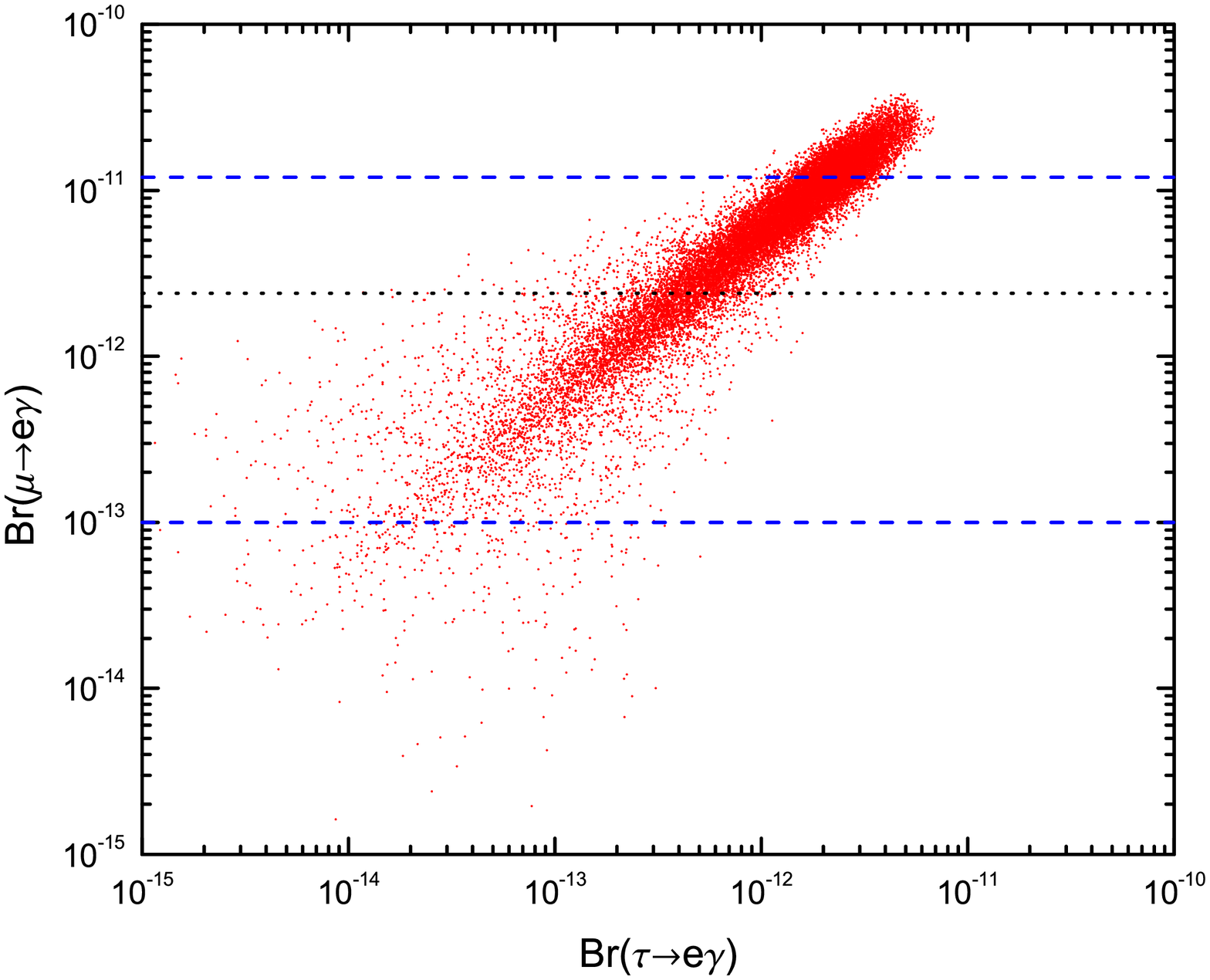}&\includegraphics[scale=.30]{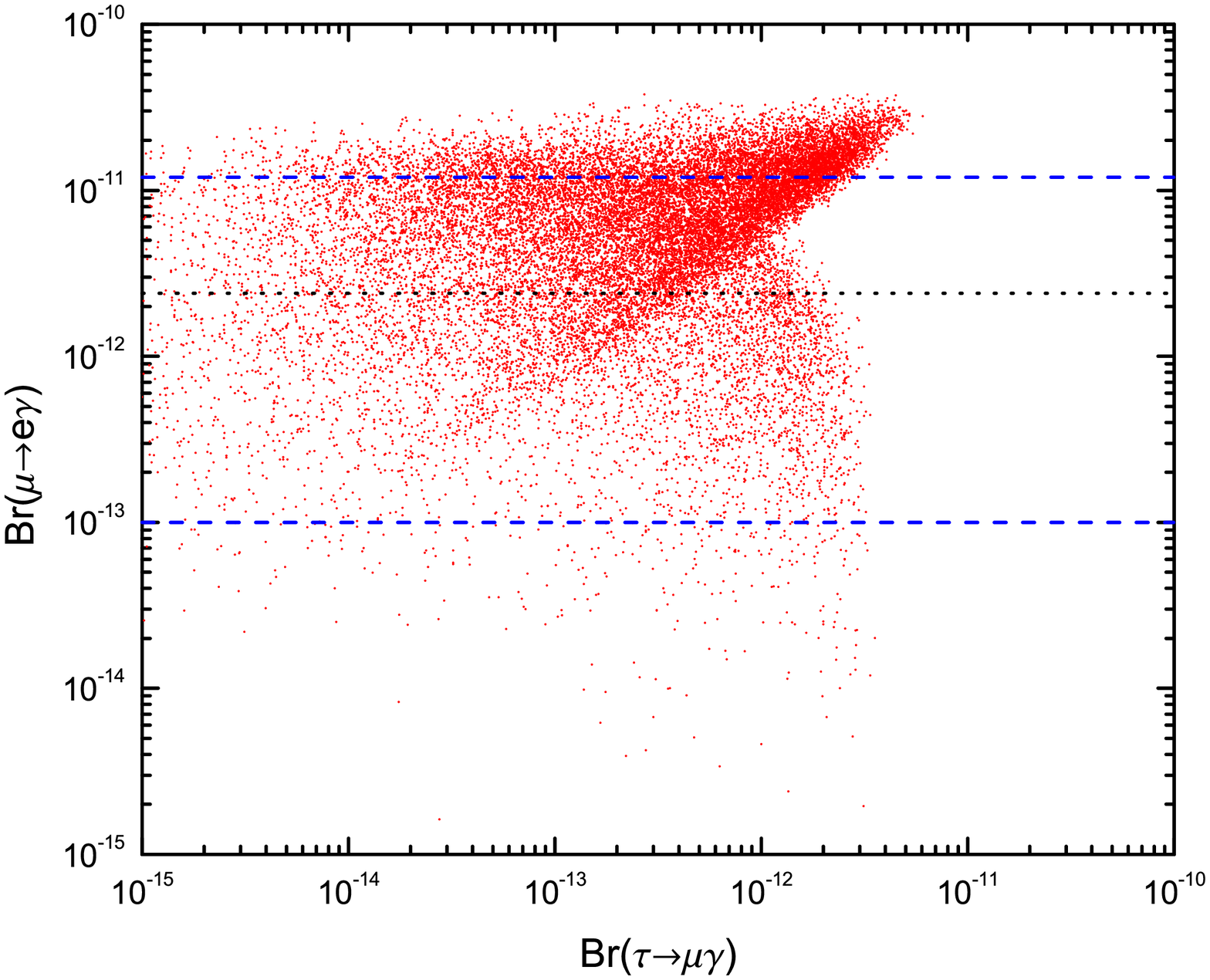}
\end{tabular}
\caption{\label{fig:LFV}\it  Correlation between the LFV branching ratios $Br(\mu\rightarrow e\gamma)$, $Br(\tau\rightarrow e\gamma)$ and $Br(\tau\rightarrow\mu\gamma)$.
The horizontal dashed lines correspond to $Br(\mu\rightarrow e\gamma)=1.2\times10^{-11}$ and $Br(\mu\rightarrow e\gamma)=10^{-13}$ ,
which are the present and future sensitivities on $Br(\mu\rightarrow e\gamma)$ listed in PDG \cite{pdg}, respectively. The horizontal dotted line is the most recent experimental upper bound $2.4\times10^{-12}$ from the MEG collaboration \cite{Adam:2011ch}.}
\end{center}
\end{figure}

The results for the LFV radiative decays $\mu\rightarrow e\gamma$, $\tau\rightarrow e\gamma$ and $\tau\rightarrow\mu\gamma$ are presented in Fig. \ref{fig:LFV}, which shows
the correlation between each two of the indicated branching ratios. In these plots, we require that the corresponding $\Delta m^2_{sol}$, $\Delta m^2_{atm}$ and the three lepton mixing angles
lie in the $3\sigma$ ranges Ref. \cite{Fogli:2011qn}. It is clear that $\tau\rightarrow e\gamma$ and $\tau\rightarrow\mu\gamma$ are far below from both the present
and future experimental sensitivities in all the parameter space; although the $Br(\mu\rightarrow e\gamma)$ is above the latest MEG upper bound in a
relevant part of the parameter space, still a sizable amount of
points are not excluded. The expected sensitivity of future experiments around $10^{-13}$ listed in PDG \cite{pdg}
will put severe constraints on the model.
All LFV branching ratios are predicted to be rather small; this distinguishing feature of our construction is mainly due to their dependence on the fourth power of the neutrino Yukawa couplings which are of ${\cal O}(\lambda^2_c)$ at LO.
In the left panel of Fig. \ref{fig:LFV}, the densely populated straight band represents the ratio
$Br(\tau\rightarrow e\gamma)/Br(\mu\rightarrow e\gamma)$ around 0.18, in agreement with Eq.(\ref{eq:LFV_relation}). The correlation between $Br(\mu\rightarrow e\gamma)$ and $Br(\tau\rightarrow\mu\gamma)$, which is shown in
the right panel of Fig. \ref{fig:LFV}, is somewhat involved and is due to the fact that they are non-linearly correlated, as it has been stated below
Eq.(\ref{eq:LFV_relation}). We have checked that the shape of the correlation
follows exactly the LO prescription. Moreover, $Br(\tau\rightarrow\mu\gamma)$ is always found to be smaller than $Br(\tau\rightarrow e\gamma)$,
in contrast with previous $A_4$
models where $Br(\tau\rightarrow\mu\gamma)$ is approximately one order of magnitude larger than $Br(\tau\rightarrow e\gamma)$ \cite{Hagedorn:2009df}. All these numerical results are
consistent with our LO analysis. For the other LFV processes, we find that the trilepton decays $\tau\rightarrow3e$ and $\tau\rightarrow3\mu$ are predicted to be about six to
eight orders of magnitude below the present and future sensitivities in all the allowed parameter space and  $\mu\rightarrow3e$ and $\mu-e$ conversion in $^{48}_{22}$Ti
and $^{27}_{13}$Al
are always below the present upper bounds\footnote{There are no plans to perform a new experimental searching for the $\mu\rightarrow3e$ decay with higher precision.}. However,
the $\mu-e$ conversion processes are within the reach of next generation experiments in a considerable part of the parameter space; in particular,  $\mu-e$ conversion in $^{48}_{22}$Ti is expected to play an important role, due to the drastic improvement of the experimental sensitivity.
The constraints imposed by future $\mu-e$ conversion experiments would be much stronger that those
from the radiative decay $\mu\rightarrow e\gamma$. If $\mu-e$ conversion in $^{48}_{22}$Ti is not observed in the future, the present model would be almost ruled out.

\section{Conclusion and discussion}

In this work we have constructed a new model for TB mixing with a global flavor symmetry $A_4\times Z_4\times Z_2$. All the right-handed matter fields
are assigned to $A_4$ singlets: the three right-handed neutrinos $\nu^{c}_i$ transform as $\mathbf{1}$, $\mathbf{1}'$ and $\mathbf{1}''$ and the right-handed charged
leptons $e^{c}$, $\mu^{c}$ and $\tau^{c}$ are all invariant under $A_4$. The three generations of leptons form an $A_4$ triplet $\mathbf{3}$ as usual.
The easiest way to break the flavor symmetry is to introduce only one flavon in the neutrino sector but the resulting neutrino mass spectrum gives
$m_2$=0 or $m_1=m_3$ and then incompatible with the data.
As a consequence, we should introduce at least two flavon fields in the neutrino sector to obtain a phenomenologically viable model.
This statement generally holds if the right-handed neutrinos are assumed to be $A_4$ singlets. The model would appear more symmetric if we let the charged lepton
fields $e^{c}$, $\mu^{c}$ and $\tau^{c}$ to transform as $\mathbf{1}$, $\mathbf{1}'$ and $\mathbf{1}''$ as well. But in this case we would need to introduce a Froggatt-Nielsen $U(1)$ symmetry to generate the charged lepton mass hierarchies. It is remarkable that we can still produce TB mixing although the $A_4$ symmetry is broken completely at LO in both the neutrino and charged lepton sectors;
the reason being an accidental $Z_2\times Z_3$ symmetry preserved in the mass matrices. This is a special feature of present model.

The light neutrino mass spectrum is predicted to be normal
hierarchy and the lightest neutrino mass is zero at LO although three right-handed neutrinos are introduced (generally massless neutrino are
realized via the minimal see-saw mechansim). The effective mass of neutrinoless double-$\beta$ decay is predicted to be about 2.9 meV, which is much below the prospective sensitivities
of future experiments. All three lepton mixing angles receive corrections of order $\lambda^2_c$ from the NLO corrections since no special dynamics is introduced to separate the corrections to $\theta_{13}$ from the other angles. Although the
reactor angle $\theta_{13}$ is expected to be of order $\lambda^2_c$, we have shown that our model can accommodate values of $\theta_{13}$ consistent with the recent experimental results. Consequently the model cannot be ruled out by present data, although there is hint of relatively larger $\theta_{13}$ from the T2K collaboration. Precise measurement of $\theta_{13}$ is the most direct test of the model: if it is found to be near the present upper bound, a large class of discrete flavor symmetry models (including the present one) would be ruled out and  the TB mixing may not be a good starting point for model building.

The predictions for leptogenesis and LFV branching ratios are analyzed in detail. We find that the first heavy right-handed neutrino does not contribute to the leptonic CP asymmetry
even if NLO corrections are taken into account. The leptogenesis is realized via the so-called resonant leptogenesis of the second and third heavy neutrinos which are degenerate at
LO. In order to account for the observed baryon asymmetry, moderate fine tuning among the neutrino Yukawa couplings is required. The LFV branching ratios are predicted to be rather small because they are proportional to the fourth power of the neutrino Yukawa couplings, of order $\lambda^2_c$ at LO. We find that the LFV processes $\mu\rightarrow e\gamma$ and $\mu-e$ conversion in $^{48}_{22}$Ti and $^{27}_{13}$Al are within the reach of next generation of experiments; in particular, $\mu-e$ conversion in $^{48}_{22}$Ti could impose extremely strong constraints on the model due to the considerable improvement of its sensitivity in near future, whereas the branching fractions of
$\tau\rightarrow e\gamma$, $\tau\rightarrow\mu\gamma$ and trilepton decay $\ell_i\rightarrow3\ell_j$ are far below the present and future sensitivities. The above theoretical predictions for neutrinoless double-$\beta$ decay and LFV processes are other important tests of the model.

\section*{Acknowledgements}

We are grateful to IFIC for their hospitality while part of this work was being completed during the FLASY meeting.
G.J.D is supported by the National Natural Science Foundation of China under Grant No.10905053, Chinese Academy KJCX2-YW-N29 and the 973 project with Grant No. 2009CB825200. D.M. acknowledges MIUR (Italy) for financial support under the contract PRIN08.

\vfill
\newpage

\section*{Appendix: The discrete group $A_4$}
In this appendix, we briefly review some basic properties of the $A_4$ group. $A_4$ is the even permutation group of four objects, it has 12 elements.
Geometrically, it is the symmetry group of a regular tetrahedron. The elements of $A_4$ can be generated by two generators $S$ and $T$ obeying the relation:
\begin{equation}
\label{eq:relation}S^2=T^3=(ST)^3=1\,.
\end{equation}
The 12 elements of $A_4$ are obtained as $1$, $S$, $T$, $ST$, $TS$, $T^2$, $ST^2$, $STS$, $TST$, $T^2S$, $TST^2$ and $T^2ST$. Without loss of generality, we can choose
\begin{equation}
S=(14)(23),~~~~~T=(123)\,,
\end{equation}
where the cycle $(123)$ represents the permutation $(1,2,3,4)\rightarrow(2,3,1,4)$ and $(14)(23)$ means $(1,2,3,4)\rightarrow(4,3,2,1)$.
The $A_4$ elements belong to 4 conjugate classes:
\begin{eqnarray}
\nonumber&&{\cal C}_1: 1\\
\nonumber&&{\cal C}_2: T=(123),~~ST=(134),~~TS=(142),~~STS=(243)\\
\nonumber&&{\cal C}_3: T^2=(132),~~ST^2=(124),~~T^2S=(143),~~ST^2S=(234)\\
&&{\cal C}_4: S=(14)(23),~~T^2ST=(12)(34),~~TST^2=(13)(24)\,.
\end{eqnarray}
There are 4 inequivalent irreducible representations of $A_4$: three singlet representation $\mathbf{1}$, $\mathbf{1}'$, $\mathbf{1}''$ and one triplet
representation $\mathbf{3}$. For the one-dimensional representations, from the generator relation in Eq.(\ref{eq:relation}), we can easily obtain
that the representations are given by:
\begin{eqnarray}
\nonumber&& 1:~~S=1,~~T=1\\
\nonumber&&1':~~S=1,~~T=\omega^2\\
\label{eq:rep1}&&1'':~~S=1,~~T=\omega\,,
\end{eqnarray}
where $\omega=e^{2\pi i/3}$ is the cube root of unit. For the three-dimensional representation, in the basis where $T$ is diagonal, it is given by
\begin{equation}
\label{eq:rep2}3:~~S=\frac{1}{3}\left(\begin{array}{ccc}
-1&2&2\\
2&-1&2\\
2&2&-1
\end{array}\right),~~~~T=\left(\begin{array}{ccc}1&0&0\\
0&\omega^2&0\\
0&0&\omega
\end{array}\right)\,.
\end{equation}
The multiplication rules between various irreducible representations are as follows:
\begin{eqnarray}
\nonumber&&1\otimes R=R,~~1'\otimes1''=1,~~~1'\otimes1'=1'',~~~1''\otimes1''=1',\\
&&3\otimes3=1\oplus1'\oplus1''\oplus3_S\oplus3_A,~~3\otimes1'=3,~~3\otimes1''=3\,,
\end{eqnarray}
where $R$ denotes any $A_4$ representation. From Eq.(\ref{eq:rep1}) and Eq.(\ref{eq:rep2}), we can straightforwardly obtain the decomposition of the product
representations. For two $A_4$ triplets $\alpha=(\alpha_1,\alpha_2,\alpha_3)$ and $\beta=(\beta_1,\beta_2,\beta_3)$, we have:
\begin{eqnarray}
\nonumber&&1\equiv(\alpha\beta)=\alpha_1\beta_1+\alpha_2\beta_3+\alpha_3\beta_2\\
\nonumber&&1'\equiv(\alpha\beta)'=\alpha_3\beta_3+\alpha_1\beta_2+\alpha_2\beta_1\\
\nonumber&&1''\equiv(\alpha\beta)''=\alpha_2\beta_2+\alpha_1\beta_3+\alpha_3\beta_1\\
\nonumber&&3_A\equiv(\alpha\beta)_{3_A}=(\alpha_2\beta_3-\alpha_3\beta_2,\alpha_1\beta_2-\alpha_2\beta_1,\alpha_3\beta_1-\alpha_1\beta_3)\\
&&3_S\equiv(\alpha\beta)_{3_S}=(2\alpha_1\beta_1-\alpha_2\beta_3-\alpha_3\beta_2,2\alpha_3\beta_3-\alpha_1\beta_2-\alpha_2\beta_1,2\alpha_2\beta_2-\alpha_1\beta_3-\alpha_3\beta_1)
\end{eqnarray}
Furthermore, if $\gamma$, $\gamma'$ and $\gamma''$ are $A_4$ singlets transforming as $\mathbf{1}$, $\mathbf{1}'$ and $\mathbf{1}''$,
then the products $\alpha\gamma$, $\alpha\gamma'$ and $\alpha\gamma''$ are triplets explicitly given by
$(\alpha_1\gamma,\alpha_2\gamma,\alpha_3\gamma)$ and $(\alpha_3\gamma',\alpha_1\gamma',\alpha_2\gamma')$ and $(\alpha_2\gamma'',\alpha_3\gamma'',\alpha_1\gamma'')$.
It is interesting to note that if the $A_4$ flavor symmetry is broken down to the $Z_2$ subgroup generated by $S$ in the neutrino sector, then the neutrino mass
matrix $m_{\nu}$ is invariant under the action of $S$, i.e., $S^{T}m_{\nu}S=m_{\nu}$; consequently $m_{\nu}$ has the general form:
\begin{equation}
m_{\nu}=\left(\begin{array}{ccc}
A&B&C\\
B&D& A+C-D\\
C& A+C-D &B-C+D
\end{array}\right)\,.
\end{equation}
It admits the eigenvector (1,1,1) so that the trimaximal mixing can be reproduced naturally.
With proper choice of flavon fields in the neutrino sector, e.g., in the absence of flavon transforming as $\mathbf{1}'$ or $\mathbf{1}''$,
the $\mu-\tau$ symmetry arises accidentally. As a consequence, we have $B=C$ and the resulting neutrino mass matrix is exactly diagonalized by the TB mixing matrix.

\end{document}